\documentclass[aps,twocolumn,prd,showpacs,showkeys,preprintnumbers,superscriptaddress,nobibnotes,floatfix,longbibliography,notitlepage,nofootinbib]{revtex4-2}

\usepackage{amsmath}
\usepackage{amsfonts}
\usepackage{amssymb}
\usepackage{graphicx}
\usepackage[colorlinks=true,allcolors=blue]{hyperref}
\usepackage{relsize}
\usepackage{gensymb}
\usepackage{fontawesome}
\usepackage[capitalize]{cleveref}
\usepackage[dvipsnames,table]{xcolor}
\usepackage{placeins}

\usepackage{multirow} % for \multirow command

\usepackage{lineno}
\begin{document}

\title{Methods and stability tests associated with the sterile neutrino search using improved high-energy $\nu_\mu$ event reconstruction in IceCube}

\affiliation{III. Physikalisches Institut, RWTH Aachen University, D-52056 Aachen, Germany}
\affiliation{Department of Physics, University of Adelaide, Adelaide, 5005, Australia}
\affiliation{Dept. of Physics and Astronomy, University of Alaska Anchorage, 3211 Providence Dr., Anchorage, AK 99508, USA}
\affiliation{Dept. of Physics, University of Texas at Arlington, 502 Yates St., Science Hall Rm 108, Box 19059, Arlington, TX 76019, USA}
\affiliation{CTSPS, Clark-Atlanta University, Atlanta, GA 30314, USA}
\affiliation{School of Physics and Center for Relativistic Astrophysics, Georgia Institute of Technology, Atlanta, GA 30332, USA}
\affiliation{Dept. of Physics, Southern University, Baton Rouge, LA 70813, USA}
\affiliation{Dept. of Physics, University of California, Berkeley, CA 94720, USA}
\affiliation{Lawrence Berkeley National Laboratory, Berkeley, CA 94720, USA}
\affiliation{Institut f{\"u}r Physik, Humboldt-Universit{\"a}t zu Berlin, D-12489 Berlin, Germany}
\affiliation{Fakult{\"a}t f{\"u}r Physik {\&} Astronomie, Ruhr-Universit{\"a}t Bochum, D-44780 Bochum, Germany}
\affiliation{Universit{\'e} Libre de Bruxelles, Science Faculty CP230, B-1050 Brussels, Belgium}
\affiliation{Vrije Universiteit Brussel (VUB), Dienst ELEM, B-1050 Brussels, Belgium}
\affiliation{Department of Physics and Laboratory for Particle Physics and Cosmology, Harvard University, Cambridge, MA 02138, USA}
\affiliation{Dept. of Physics, Massachusetts Institute of Technology, Cambridge, MA 02139, USA}
\affiliation{Dept. of Physics and The International Center for Hadron Astrophysics, Chiba University, Chiba 263-8522, Japan}
\affiliation{Department of Physics, Loyola University Chicago, Chicago, IL 60660, USA}
\affiliation{Dept. of Physics and Astronomy, University of Canterbury, Private Bag 4800, Christchurch, New Zealand}
\affiliation{Dept. of Physics, University of Maryland, College Park, MD 20742, USA}
\affiliation{Dept. of Astronomy, Ohio State University, Columbus, OH 43210, USA}
\affiliation{Dept. of Physics and Center for Cosmology and Astro-Particle Physics, Ohio State University, Columbus, OH 43210, USA}
\affiliation{Niels Bohr Institute, University of Copenhagen, DK-2100 Copenhagen, Denmark}
\affiliation{Dept. of Physics, TU Dortmund University, D-44221 Dortmund, Germany}
\affiliation{Dept. of Physics and Astronomy, Michigan State University, East Lansing, MI 48824, USA}
\affiliation{Dept. of Physics, University of Alberta, Edmonton, Alberta, T6G 2E1, Canada}
\affiliation{Erlangen Centre for Astroparticle Physics, Friedrich-Alexander-Universit{\"a}t Erlangen-N{\"u}rnberg, D-91058 Erlangen, Germany}
\affiliation{Physik-department, Technische Universit{\"a}t M{\"u}nchen, D-85748 Garching, Germany}
\affiliation{D{\'e}partement de physique nucl{\'e}aire et corpusculaire, Universit{\'e} de Gen{\`e}ve, CH-1211 Gen{\`e}ve, Switzerland}
\affiliation{Dept. of Physics and Astronomy, University of Gent, B-9000 Gent, Belgium}
\affiliation{Dept. of Physics and Astronomy, University of California, Irvine, CA 92697, USA}
\affiliation{Karlsruhe Institute of Technology, Institute for Astroparticle Physics, D-76021 Karlsruhe, Germany}
\affiliation{Karlsruhe Institute of Technology, Institute of Experimental Particle Physics, D-76021 Karlsruhe, Germany}
\affiliation{Dept. of Physics, Engineering Physics, and Astronomy, Queen's University, Kingston, ON K7L 3N6, Canada}
\affiliation{Department of Physics {\&} Astronomy, University of Nevada, Las Vegas, NV 89154, USA}
\affiliation{Nevada Center for Astrophysics, University of Nevada, Las Vegas, NV 89154, USA}
\affiliation{Dept. of Physics and Astronomy, University of Kansas, Lawrence, KS 66045, USA}
\affiliation{Centre for Cosmology, Particle Physics and Phenomenology - CP3, Universit{\'e} catholique de Louvain, Louvain-la-Neuve, Belgium}
\affiliation{Department of Physics, Mercer University, Macon, GA 31207-0001, USA}
\affiliation{Dept. of Astronomy, University of Wisconsin{\textemdash}Madison, Madison, WI 53706, USA}
\affiliation{Dept. of Physics and Wisconsin IceCube Particle Astrophysics Center, University of Wisconsin{\textemdash}Madison, Madison, WI 53706, USA}
\affiliation{Institute of Physics, University of Mainz, Staudinger Weg 7, D-55099 Mainz, Germany}
\affiliation{Department of Physics, Marquette University, Milwaukee, WI 53201, USA}
\affiliation{Institut f{\"u}r Kernphysik, Westf{\"a}lische Wilhelms-Universit{\"a}t M{\"u}nster, D-48149 M{\"u}nster, Germany}
\affiliation{Bartol Research Institute and Dept. of Physics and Astronomy, University of Delaware, Newark, DE 19716, USA}
\affiliation{Dept. of Physics, Yale University, New Haven, CT 06520, USA}
\affiliation{Columbia Astrophysics and Nevis Laboratories, Columbia University, New York, NY 10027, USA}
\affiliation{Dept. of Physics, University of Oxford, Parks Road, Oxford OX1 3PU, United Kingdom}
\affiliation{Dipartimento di Fisica e Astronomia Galileo Galilei, Universit{\`a} Degli Studi di Padova, I-35122 Padova PD, Italy}
\affiliation{Dept. of Physics, Drexel University, 3141 Chestnut Street, Philadelphia, PA 19104, USA}
\affiliation{Physics Department, South Dakota School of Mines and Technology, Rapid City, SD 57701, USA}
\affiliation{Dept. of Physics, University of Wisconsin, River Falls, WI 54022, USA}
\affiliation{Dept. of Physics and Astronomy, University of Rochester, Rochester, NY 14627, USA}
\affiliation{Department of Physics and Astronomy, University of Utah, Salt Lake City, UT 84112, USA}
\affiliation{Dept. of Physics, Chung-Ang University, Seoul 06974, Republic of Korea}
\affiliation{Oskar Klein Centre and Dept. of Physics, Stockholm University, SE-10691 Stockholm, Sweden}
\affiliation{Dept. of Physics and Astronomy, Stony Brook University, Stony Brook, NY 11794-3800, USA}
\affiliation{Dept. of Physics, Sungkyunkwan University, Suwon 16419, Republic of Korea}
\affiliation{Institute of Basic Science, Sungkyunkwan University, Suwon 16419, Republic of Korea}
\affiliation{Institute of Physics, Academia Sinica, Taipei, 11529, Taiwan}
\affiliation{Dept. of Physics and Astronomy, University of Alabama, Tuscaloosa, AL 35487, USA}
\affiliation{Dept. of Astronomy and Astrophysics, Pennsylvania State University, University Park, PA 16802, USA}
\affiliation{Dept. of Physics, Pennsylvania State University, University Park, PA 16802, USA}
\affiliation{Dept. of Physics and Astronomy, Uppsala University, Box 516, SE-75120 Uppsala, Sweden}
\affiliation{Dept. of Physics, University of Wuppertal, D-42119 Wuppertal, Germany}
\affiliation{Deutsches Elektronen-Synchrotron DESY, Platanenallee 6, D-15738 Zeuthen, Germany}

\author{R. Abbasi}
\affiliation{Department of Physics, Loyola University Chicago, Chicago, IL 60660, USA}
\author{M. Ackermann}
\affiliation{Deutsches Elektronen-Synchrotron DESY, Platanenallee 6, D-15738 Zeuthen, Germany}
\author{J. Adams}
\affiliation{Dept. of Physics and Astronomy, University of Canterbury, Private Bag 4800, Christchurch, New Zealand}
\author{S. K. Agarwalla}
\thanks{also at Institute of Physics, Sachivalaya Marg, Sainik School Post, Bhubaneswar 751005, India}
\affiliation{Dept. of Physics and Wisconsin IceCube Particle Astrophysics Center, University of Wisconsin{\textemdash}Madison, Madison, WI 53706, USA}
\author{J. A. Aguilar}
\affiliation{Universit{\'e} Libre de Bruxelles, Science Faculty CP230, B-1050 Brussels, Belgium}
\author{M. Ahlers}
\affiliation{Niels Bohr Institute, University of Copenhagen, DK-2100 Copenhagen, Denmark}
\author{J.M. Alameddine}
\affiliation{Dept. of Physics, TU Dortmund University, D-44221 Dortmund, Germany}
\author{N. M. Amin}
\affiliation{Bartol Research Institute and Dept. of Physics and Astronomy, University of Delaware, Newark, DE 19716, USA}
\author{K. Andeen}
\affiliation{Department of Physics, Marquette University, Milwaukee, WI 53201, USA}
\author{C. Arg{\"u}elles}
\affiliation{Department of Physics and Laboratory for Particle Physics and Cosmology, Harvard University, Cambridge, MA 02138, USA}
% ...

\collaboration{IceCube Collaboration}
\thanks{analysis@icecube.wisc.edu}
\noaffiliation

\begin{abstract}
We provide supporting details for the search for a 3+1 sterile neutrino using data collected over 10.7 years at the IceCube Neutrino Observatory.
The analysis uses atmospheric muon-flavored neutrinos from 0.5 to 100\, TeV that traverse the Earth to reach the IceCube detector, and finds a best-fit point at $\sin^2(2\theta_{24}) = 0.16$ and $\Delta m^{2}_{41} = 3.5$~eV$^2$ with a goodness-of-fit p-value of 12\% and consistency with the null hypothesis of no oscillations to sterile neutrinos with a p-value of 3.1\%.
Several improvements were made over past analyses, which are reviewed in this article, including upgrades to the reconstruction and the study of sources of systematic uncertainty.
We provide details of the fit quality and discuss stability tests that split the data for separate samples, comparing results.
We find that the fits are consistent between split data sets.
\end{abstract}

\maketitle

\section{Introduction}

Longstanding anomalies observed in accelerator-, reactor- and source-based neutrino experiments~\cite{Diaz:2019fwt,Dentler:2018sju,Gariazzo:2017fdh} have motivated the exploration of new physics models that introduce an additional mass state along with a new non-interacting, or ``sterile,'' neutrino flavor.
As in the three-flavor model, the mass states are rotated with respect to the flavor states by a 4$\times$4 matrix that includes the Pontecorvo-Maki-Nakagawa-Sakata (PMNS) matrix within it. 
Full consideration of such a model, often called ``3+1,'' in an oscillation framework leads to additional measurable parameters: a mass-squared splitting, $\Delta m^2_{41}$, three elements of the mixing matrix that are often expressed as angles:  $\theta_{14}$, $\theta_{24}$, and $\theta_{34}$; and $CP$-phases: $\delta_{14}$ and $\delta_{24}$.

The accessible neutrino flavors that we use to test this model, through searches for vacuum oscillations, are $\nu_e$ and $\nu_\mu$.
The LSND~\cite{PhysRevLett.77.3082} and MiniBooNE~\cite{PhysRevLett.110.161801} experiments have reported evidence for $\nu_{\mu}\rightarrow\nu_e$ (called ``$\nu_e$ appearance'') and the BEST experiment~\cite{PhysRevLett.128.232501} has reported evidence for $\nu_e \rightarrow \nu_e$ (``$\nu_e$ disappearance''). 
Both sets of results are consistent with $\Delta m^2_{41}$ between 1 and 10\,eV$^2$.
However, there is no evidence for $\nu_\mu \rightarrow \nu_\mu$ (``$\nu_\mu$ disappearance'') at this mass splitting, even though this is a necessary feature of the 3+1 model.

This tension has motivated a recent program of searches for high-energy $\nu_\mu$ disappearance using data from the IceCube Neutrino Observatory located in the Antarctic ice at the South Pole.
The 3+1 neutrino model can manifest in two distinct ways both dependent on mass splitting and mixing angles: vacuum oscillations and matter-enhanced resonant transitions in antineutrino flavors as they traverse the Earth's dense core.
IceCube provides a large data set of high-energy atmospheric and astrophysical $\nu_\mu$ traversing the Earth with energies ranging from several hundreds of GeV to a few PeV.
An eV-scale sterile state would distort the observed neutrino flux due to the interplay of both vacuum oscillations and matter-enhanced resonances.
Hence, an analysis of IceCube data allows us to examine the two manifestations of 3+1 effects simultaneously, representing a more robust test than searching for a single oscillation signature.

\begin{figure}[tb!]
    \centering
    \includegraphics[width=0.5\textwidth]{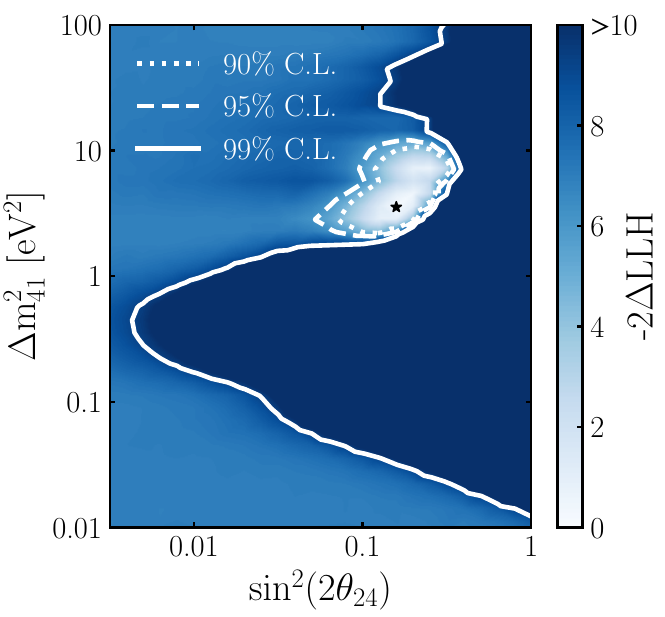}
    \caption{\textbf{Result of the IceCube search for $\nu_\mu$ disappearance with 10.7 years of data}.  The color scale shows the likelihood difference relative to the best-fit point.
    The best-fit point is marked with a black star, and the 90\% (dotted), 95\% (dashed), and 99\% (solid) confidence level (C.L.) contours are drawn, assuming Wilks’ theorem with two degrees of freedom.}
    \label{fig:likelihood}
\end{figure}

The IceCube Collaboration has published results from a $\nu_\mu$ disappearance search within a 3+1 model using 10.7 years of IceCube data in Ref.~\cite{PRL}.
The analysis found a best-fit point at $\sin^2(2\theta_{24}) = 0.16$ and $\Delta m^{2}_{41} = 3.5$ eV$^2$ with a goodness-of-fit p-value of 12\% and consistency with the no sterile neutrino (``null'') model given by a of p-value of 3.1\%.
Fig.~\ref{fig:likelihood} shows the result interpreted using Wilks' theorem, assuming two degrees of freedom.
Given this consistency with the null hypothesis, the result does not represent evidence for $\nu_\mu$ disappearance, but it provides valuable information for our understanding of the 3+1 neutrino landscape.

This article provides a detailed look at the IceCube 3+1 analysis.
Information on the new reconstruction and event selection is provided.
Also, a series of studies of the stability of the result, explored after the result was unblinded, is reported.

\section{The Data Sample}

This analysis uses charged-current (CC) interactions produced in the Antarctic ice by the up-going $\nu_\mu$ flux, that is, the neutrinos that arrive at the detector from below the horizon.
Events may be produced outside the detector with a muon that traverses the active region (``through-going events'') or may be produced within the detector with a muon that exits the active region (``starting events'').
This analysis uses the atmospheric and astrophysical $\nu_\mu$ flux between 0.5 and 100\,TeV.

We begin by describing events in the IceCube detector, followed by discussing the performance of the reconstruction package developed for this analysis, and end by providing the event selection criteria for this analysis, resulting in a dataset of 368,071 events collected from May 13th, 2011 to June 7th, 2022.

\subsection{Events in the IceCube Detector}

The IceCube Neutrino Observatory~\cite{Aartsen:2016nxy} is a one-gigaton ice Cherenkov detector consisting of 5160 digital optical modules (DOMs)~\cite{IceCube:2008qbc} arranged on 86 strings, located between 1450~m and 2450~m below the top of the Antarctic ice in a hexagonal pattern.
Most of the detector has a sparse spacing of 17~m between DOMs on a string, and $\sim$125~m laterally between strings, which leads to an energy threshold of $\sim 100$\,GeV. 
At the bottom center of the detector, there is a more densely instrumented 8-string region with a 7\,m vertical spacing between DOMs and $\sim 50$\,m between strings, called ``DeepCore,'' that has a $\sim 10$\,GeV threshold~\cite{IceCube:2011ucd}.  

Events in this analysis are generated by interactions of high-energy atmospheric and astrophysical neutrinos.
Atmospheric neutrinos arise from the interactions of primary cosmic rays with nuclei in the Earth’s atmosphere. 
Neutrinos from pions and kaons are called the ``conventional flux,'' and the rates are well-predicted due to an extensive campaign at accelerators that includes the NA49 and NA61 experiments at CERN, though measurements by HARP, PHENIX and STAR are relevant in the lower and higher energy bands~\cite{Barr:2006it, Fedynitch:2022vty}.
This analysis uses the DAEMONFLUX model, which is a recent description of the conventional flux~\cite{Yanez:2023lsy}.
Neutrinos produced by prompt interactions, such as from charm decays, as well as by astrophysical sources make up the non-conventional flux that is modeled in the analysis through a broken power law.

The detector observes Cherenkov light deposited in the ice by the charged particles produced by the $\nu_\mu$ CC interactions near or within the detector, which are simulated using \texttt{LeptonInjector}~\cite{IceCube:2020tcq}
The light is detected by DOMs that consist of 10-inch photomultiplier tubes (PMTs)~\cite{Abbasi:2008aa} installed facing downward and the associated electronics for triggering and read-out.
From the recorded PMT waveforms, time and charge information are extracted for use in event reconstruction.

Measurement of absorption and scattering of light in the ice is crucial to event reconstruction.
At depths below 1450\,m, bubbles no longer play a significant role in the effective scattering or absorption lengths of 400\,nm light. 
Depending on the depth, the effective scattering length ranges from 20\,m to 80\,m, and the absorption length ranges from 100\,m to 400\,m.
\emph{In situ} LED light sources allow calibration of the light observed by DOMs showing that layers of glaciologically contemporaneous ice are not perfectly horizontal but vary in depth across the array, usually referred to as bulk ice tilt.
This is incorporated in the ice model~\cite{Rongen:2019wsh}, which is used to model photon propagation for both simulation and reconstruction.
Because the DOMs are inserted into a melted column of ice that re-freezes, the ``hole ice'' differs from the bulk ice, particularly due to bubbles (introduced during the hole-drilling process), and this effect is incorporated into the ice model. 

\subsection{Improved Reconstruction Algorithms} 

Vacuum oscillations depend upon two experimental parameters, neutrino energy ($E_\nu$) and the distance from where the neutrino is produced to the detection point ($L$).
IceCube's wide energy range and relatively poor resolution compared to accelerator-based experiments lead us to use fits in  $\log_{10}(E_\nu)$.
The zenith angle, $\theta_z^{\nu}$ serves as a proxy for $L$.
The samples, which will be divided into starting and through-going subsamples, will be described in the $\log_{10}(E_\nu)$ versus $\cos(\theta_z^{\nu})$ plane. Up-going neutrinos correspond to those in the zenith angle range $-1<\cos(\theta_z^{\nu})<0$.

\subsubsection{Atmospheric muon background rejection} 

Atmospheric muons are an important source of background, especially for the through-going analysis that seeks to isolate single high-energy muon tracks traversing the detector. 
In fact, despite the presence of 1.5~km of overburden above IceCube, the detector registers a trigger rate of roughly 3~kHz due to downward-going muons generated within cosmic-ray air showers.
Restricting the analysis to $\cos(\theta_z^{\mathrm{reco}})<0$ removes most cosmic-ray muons from the sample, leveraging shielding provided by the Earth.

The trajectory of each event is reconstructed using several timing-based algorithms and remains unchanged with respect to previous analysis~\cite{IceCube:2020phf}. 
Initially, a least-squares linear regression is applied to the timing distribution of the first photon observed on each DOM~\cite{AMANDA:2003vtt,Aartsen:2013bfa}.
This serves as a seed for a likelihood estimation which incorporates all the detected photons and more complex modeling factors such as the Cherenkov emission profile, ice scattering and absorption characteristics.

In addition to the up-going requirement, a set of precuts is implemented to reduce data volume and reject low-quality event candidates. 
These cuts are applied to the count and spatial distribution of triggered DOMs along the reconstructed track, as well as the length of the reconstructed track (further details are available in~\cite{IceCube:2020tka}). 
After applying these cuts, the event rate is reduced to 0.04~Hz.
However, the dataset remains predominantly populated by atmospheric muons in which the direction of the muon is mis-reconstructed.

\begin{figure}[t!]
    \centering
    \includegraphics[width=0.5\textwidth]{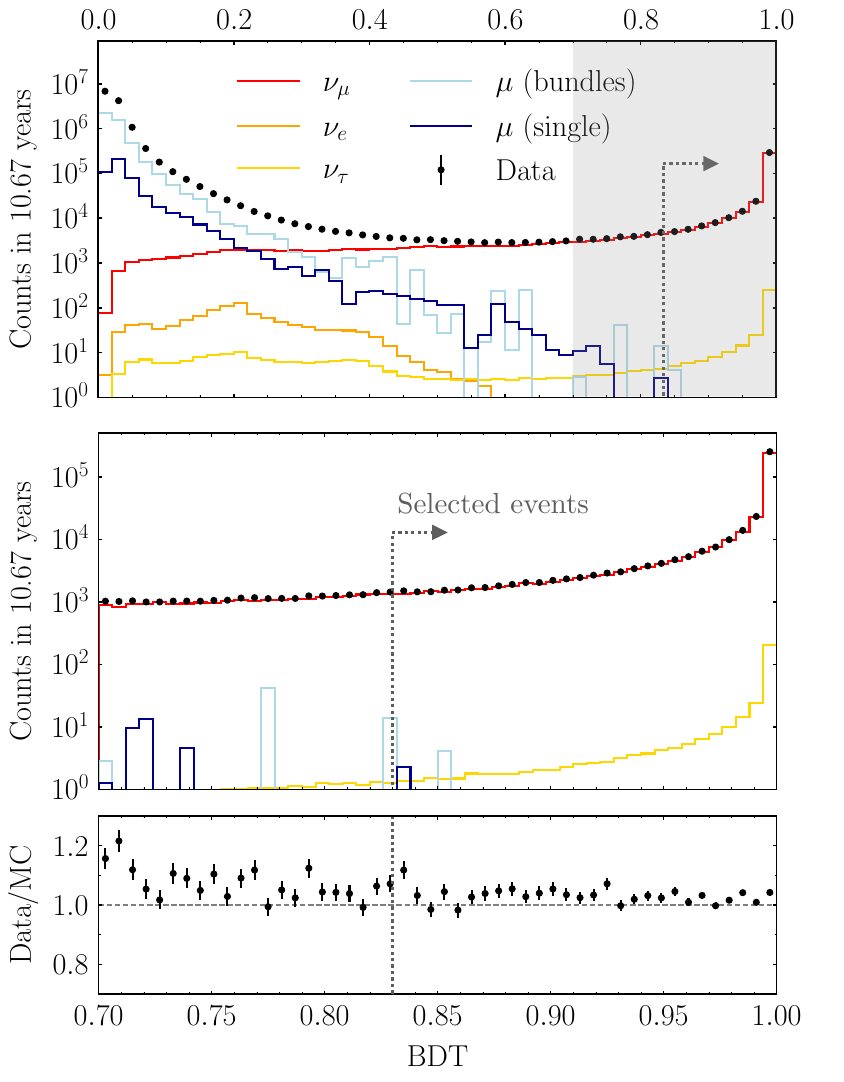}
    \caption{\textbf{BDT classifier}. The observed distribution of the BDT score is compared to the simulation under the null hypothesis and nominal systematics. 
    Data points are represented by black markers with error bars denoting statistical errors. 
    Each line corresponds to a different simulated component. 
    The upper panel displays the overall score distribution (see x-axis labels at the top), while the bottom panels zoom in on the shaded region between 0.7 and 1. 
    A vertical dashed line with an arrow indicates the minimum score for event selection. 
    Atmospheric muons that penetrate the ice overburden to reach the detector are modeled as single track (single) and multiple track (bundles) events.}
    \label{fig:bdt}
\end{figure}

In the previous analysis, we achieved 99.91\% purity of $\nu_\mu$ CC events.
In this analysis, we have taken muon rejection a step further to achieve similar purity while simultaneously increasing signal efficiency. The new step to mitigate the contamination from atmospheric muons implements a Boosted Decision Tree (BDT) using the \texttt{AdaBoostClassifier} algorithm~\cite{Hastie2009MulticlassA}.
The BDT was trained to differentiate between atmospheric muons and $\nu_\mu$ charged current interactions using 19 reconstructed observables (which were employed in the event selection of the previous analysis~\cite{IceCube:2020tka}).
Although the BDT was trained under the null hypothesis of no sterile neutrinos, we found that the BDT score distribution remained stable across different sterile neutrino hypotheses.
Fig.~\ref{fig:bdt} demonstrates the effectiveness of the BDT in distinguishing atmospheric muons from muon neutrinos.
Applying a BDT-score cut at 0.83 and using the reconstructed energy and zenith boundaries of the prior analyses\footnote{The boundaries in the previous analysis were $-1<\cos(\theta_{z}^{\mathrm{reco}})<0$ and $0.5<E_{\mathrm{reco}}/\mathrm{TeV}<10$, where $E_{\mathrm{reco}}$ was the likelihood-based estimator described in Sec.~\ref{sec:energy}.}, we achieved a nearly twofold increase in signal efficiency and a 50\% reduction in atmospheric muon background, reaching a 99.93\% purity of $\nu_\mu$ CC events.
The agreement between data and simulation within the selected region supports the estimated purity.

\subsubsection{Energy estimation}
\label{sec:energy}
The energy reconstruction employed in past sterile neutrino analyses in the $\sim 1$ TeV range estimated the neutrino energy by fitting the expected response from a template muon to the observed light deposition~\cite{Aartsen:2013vja}.
This technique exhibits limited performance for starting events, as the fitting template does not include the hadronic component of the interaction.
Thus, a primary goal of the analysis presented here has been to improve the energy reconstruction for those events.

The energy estimator used in this analysis uses the convolutional neural network (CNN) model architecture employed in previous analyses from IceCube~\cite{Abbasi:2021ryj,IceCube:2023ame}.
Charge and timing statistics of the pulses unfolded from the DOM waveforms are used as inputs to the CNN.
The pulse time series is not directly used in the reconstruction, but rather, summary variables of the time distribution are used (i.e., overall charge, width of the pulse, time of first hit, etc.).
The CNN model was trained on simulated $\nu_\mu$ CC interactions.
The label it is trained on is the ``visible energy'' of each event, which is defined as follows.
For muon neutrino interactions outside the instrumented volume, the visible energy is the muon's energy when it enters the detector.
For muon neutrino interactions occurring inside the instrumented volume, the visible energy is the muon's energy plus the scaled energy of the hadronic shower\footnote{The Cherenkov light yield for hadrons is extracted using light emission templates based on \texttt{GEANT4} simulation~\cite{Radel:2012kw}.}. 
Therefore, the visible energy closely matches the true neutrino energy for starting events.

\begin{figure}[t!]
    \centering
    \includegraphics[width=0.5\textwidth]{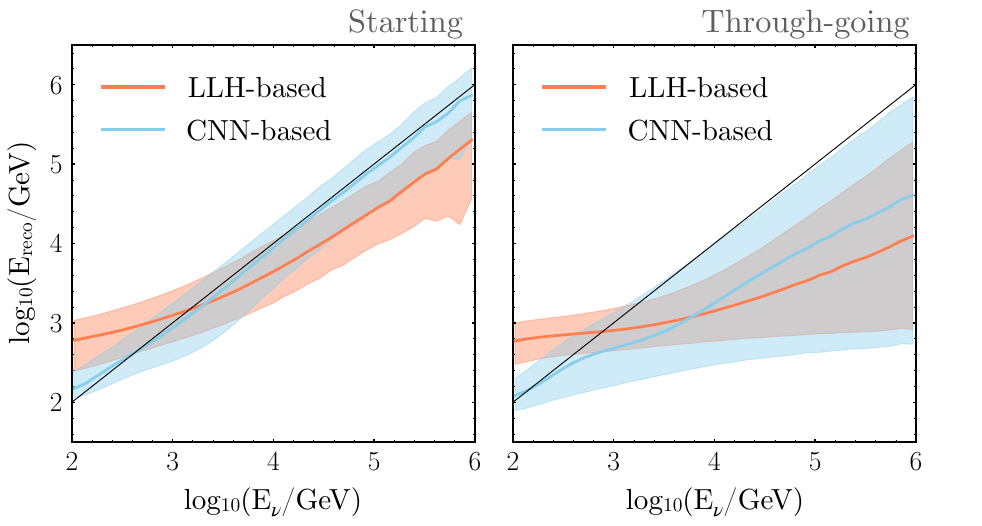}
    \caption{\textbf{Energy estimator}. Energy resolution using the LLH(likelihood)- and CNN-based energy estimators. Separate predictions are shown for events selected as starting (left) and through-going (right) as determined by the starting classifier (described in Sec.~\ref{sec:starting}). Solid orange and blue lines represent the medians of the likelihood- and CNN-based estimators, respectively. The black line shows $E_{\mathrm{reco}}/E_{\nu}=1$. Bands represent the 5th and 95th percentiles.}
    \label{fig:energy}
\end{figure}

Fig.~\ref{fig:energy} compares the energy resolution from the likelihood-based energy reconstruction used in past analyses to the CNN energy estimator used in this analysis. 
The CNN energy estimator is far more linear at low energies for both starting and through-going events, representing an important improvement.
At higher energies, the reconstruction of through-going events underestimates the neutrino energy due to energy deposition occurring outside the detector.
Consequently, the reconstructed neutrino energy resolution of through-going events is only modestly improved with the new reconstruction methods because it is dominated by the unknown hadronic energy from the interaction and the unknown energy loss as the muon traverses material before reaching the detector.
However, for the starting events, including the hadronic energy at the interaction vertex significantly improves the energy resolution for the CNN-based compared to likelihood-based reconstruction.

\subsubsection{Starting Event Identification}
\label{sec:starting}

\begin{figure}[tb!]
    \centering
    \includegraphics[width=0.5\textwidth]{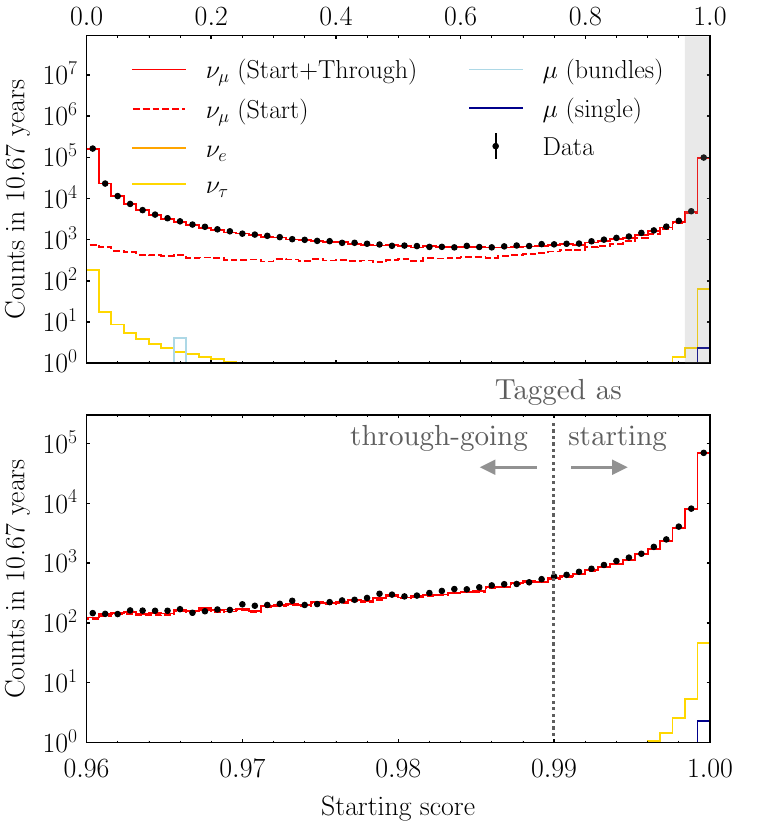}
    \caption{\textbf{Vertex classifiers}. The observed distribution of the \textit{starting score} is compared to the simulation under the null hypothesis and nominal systematics. 
    Data points are represented by black markers with error bars denoting statistical errors. 
    Each line corresponds to a different simulated component. 
    The dashed line represents true muon-neutrino CC interactions with their vertex inside the instrumented volume of IC86.
    The top panel displays the overall distribution (see x-axis labels at the top), while the bottom panels zoom in on the shaded region between 0.96 and 1.
    The vertical dashed line indicates the score that we use to classify events as either starting or through-going.
    }
    \label{fig:vertex}
\end{figure}

To discern between starting and through-going events, we employ a neural-network-based algorithm specifically designed to classify various event morphologies~\cite{Kronmueller:2019jzh}.
The distribution of starting track scores for events passing the BDT cut and within the energy-zenith boundaries of this analysis is illustrated in Fig.~\ref{fig:vertex}.
Notably, $\nu_\mu$ charged current interactions within the instrumented volume, so-called starting events, cluster at high score values. 
In this analysis, we employ a classification threshold of 0.99 to distinguish between starting and through-going events.
This threshold was chosen to obtain a very pure sample of starting events, ensuring good energy resolution within that sample. 
This choice results in 75\% of the events in the data sample being identified as through-going and 25\% as starting. 
The contamination of true through-going events in the starting event sample is 0.2\%. Conversely, the contamination of true starting events in the through-going event sample is 13\%.

\subsection{Selected Events} 
The observed and predicted number of events in the starting and through-going samples are listed in Table~\ref{tab:events}.

\begin{table}[h!]
\footnotesize
\centering
\begin{tabular}{l|r|r}
Component & \hspace*{10mm} Starting & \hspace*{1.5mm} Through-going \\
\hline
Conv. $\nu_\mu$     & 90,757.1 & 260,692.9\\
Non-conv. $\nu_\mu$ &   709.8 &   4,000.2 \\
All $\nu_e$         &     0.5 &      0.2 \\
All $\nu_\tau$      &    60.0 &    258.0 \\
Atmospheric $\mu$   &     2.3 &      4.2 \\
\hline
Total MC            & 91,529.7 & 264,955.5 \\
\hline
\hline
Data                & 93,762 & 274,309 \\
\end{tabular}
\caption{\textbf{Number of events:} The predicted and observed number of events in the starting and through-going samples collected over a livetime of 10.7 years. For the prediction, we assumed the null hypothesis using the nominal value for the systematics.}
\label{tab:events}
\end{table}

\begin{figure}[h!]
    \centering
    \includegraphics[width=0.5\textwidth]{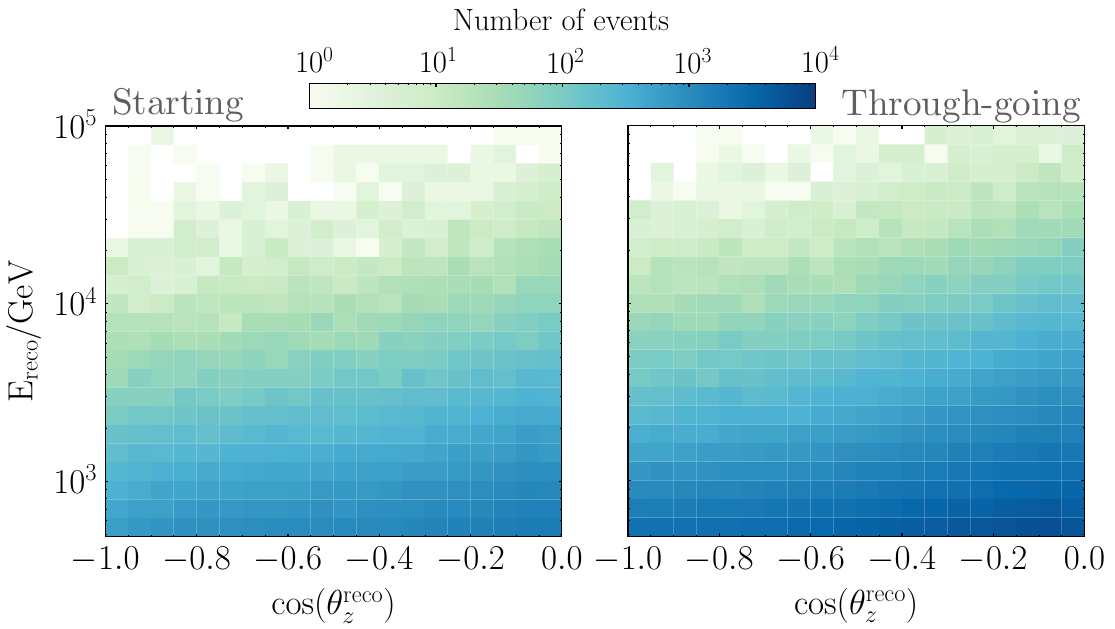}
    \caption{\textbf{Distribution of events in energy and zenith}. Number of observed events per bin in the full dataset used in this work distinguishing between starting (left) and through-going (right) events.}
    \label{fig:data_2d}
\end{figure}

Fig.~\ref{fig:data_2d} presents the observed number of events that meet the selection criteria for the starting and through-going samples after 10.7 years of data collection. 
The binning aligns with the choice used in the analysis fit. 
Both samples exhibit comparable distributions, albeit with a threefold higher rate observed in the through-going sample, attributed to its larger effective volume.
The number of events decreases with higher energy, because the neutrino fluxes (both atmospheric and astrophysical) are falling power laws. 
In addition, high energy events are depleted at low $\cos(\theta_z^{\mathrm{reco}})$ due to the intervening highly dense regions of the Earth’s core and mantle.

\section{Systematic Uncertainties}

The analysis has six broad categories of systematic uncertainties: conventional and non-conventional neutrino fluxes, bulk ice properties, local response of the DOMs, neutrino attenuation, and normalization.
In the following, we describe each of them.

\subsection{Conventional flux}
\label{sec:conv}

In the TeV regime, the primary source of muon neutrinos in IceCube comes from the decay of kaons and pions produced in cosmic-ray interactions in the atmosphere.
The uncertainty associated with this flux is broken down into components related to the modeling of hadron production, the cosmic-ray spectrum, atmospheric density, and the rate of meson energy loss in air.
We maintain the implementation of the latter two factors as in the previous analysis~\cite{IceCube:2020tka}, incorporating two nuisance parameters.

However, substantial changes have been introduced in the modeling of hadron production and the cosmic-ray spectrum. 
The previous analysis used a spectral shift parameter and ad-hoc parametrization based on the error estimates from Ref.~\cite{Barr:2006it} to model the cosmic-ray spectrum and hadronic yields, respectively . 
In contrast, the current analysis employs the \texttt{DAEMONFLUX} calculation~\cite{Yanez:2023lsy} to model this component and its associated uncertainties.

\begin{figure}[t!]
    \centering
    \includegraphics[width=0.45\textwidth]{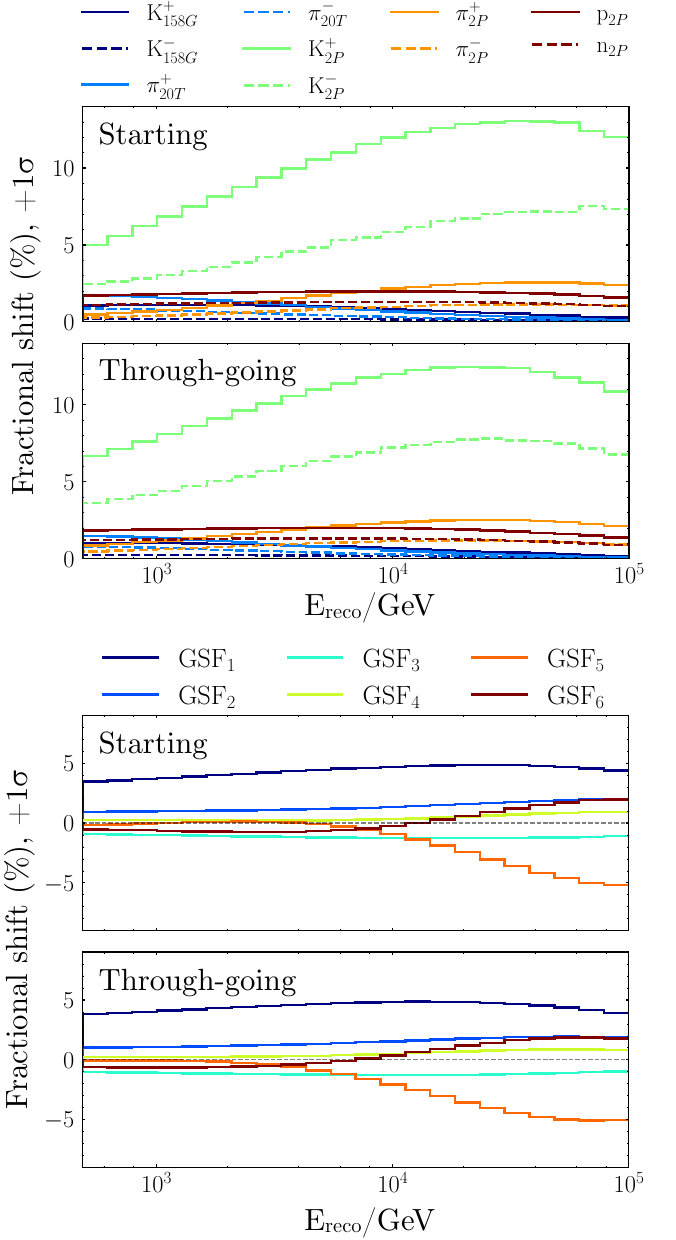}
    \caption{ \textbf{Conventional atmospheric neutrino flux systematics}. The fractional differences (in percent) in the expected number of events as a function of reconstructed energy when shifting by +1$\sigma$ each nuisance parameter associated with the hadronic yields (top panels; 10 parameters in total) and cosmic-ray spectrum (bottom panels; 6 parameters in total). In the top panels, solid (dashed) lines represent shifts in hadronic yield parameters associated with positively (negatively, and neutrons) charged particles. More information about these parameters can be found in Ref.~\cite{Yanez:2023lsy}.}
    \label{fig:grad}
\end{figure}

\texttt{DAEMONFLUX} includes parameters that characterize the hadronic yield and the cosmic-ray spectrum. 
These parameters were tuned using a combination of muon data and constraints from fixed-target experiments, yielding a model with a self-consistent method of adjusting neutrino fluxes through correlated parameters for the first time.
The reported uncertainties were leveraged to assess the impact of each parameter on our phase space.
Our analysis identifies 10 parameters associated with hadronic yields and 6 with the cosmic-ray spectrum as relevant. 
Consequently, we introduce 16 nuisance parameters with correlated Gaussian priors, adhering to the guidelines provided by the \texttt{DAEMONFLUX} calculation.
While the analysis considers the effect in the reconstructed energy-zenith space, the impact remains relatively consistent across different zenith angles.
Figure~\ref{fig:grad} illustrates the influence of each of these parameters in the reconstructed energy distribution.

The main difference compared to the previous parametrization is observed in the $\nu_\mu/\bar{\nu}_\mu$ ratio, where \texttt{DAEMONFLUX} shows larger $\nu_\mu$ contributions above 100 GeV and smaller uncertainties. Additionally, in the TeV regime, \texttt{DAEMONFLUX} and the previous model predict similar $\nu_\mu+\bar{\nu}_\mu$ fluxes (more details about these comparisons can be found in Ref.~\cite{Yanez:2023lsy}).

We computed the sensitivity to a sterile neutrino using both models, as well as performed mismodeling tests by injecting and fitting with different parametrizations. In both cases, we observed that the most conservative result was obtained when fitting with \texttt{DAEMONFLUX}. The primary reason is that \texttt{DAEMONFLUX} has more flexibility in changing the shape of the spectrum, which is critical when looking for sterile-like distortions in the flux.

\subsection{Non-conventional flux}
\label{sec:nonconv}

\begin{figure}[tb!]
\centering
\includegraphics[width=0.5\textwidth]{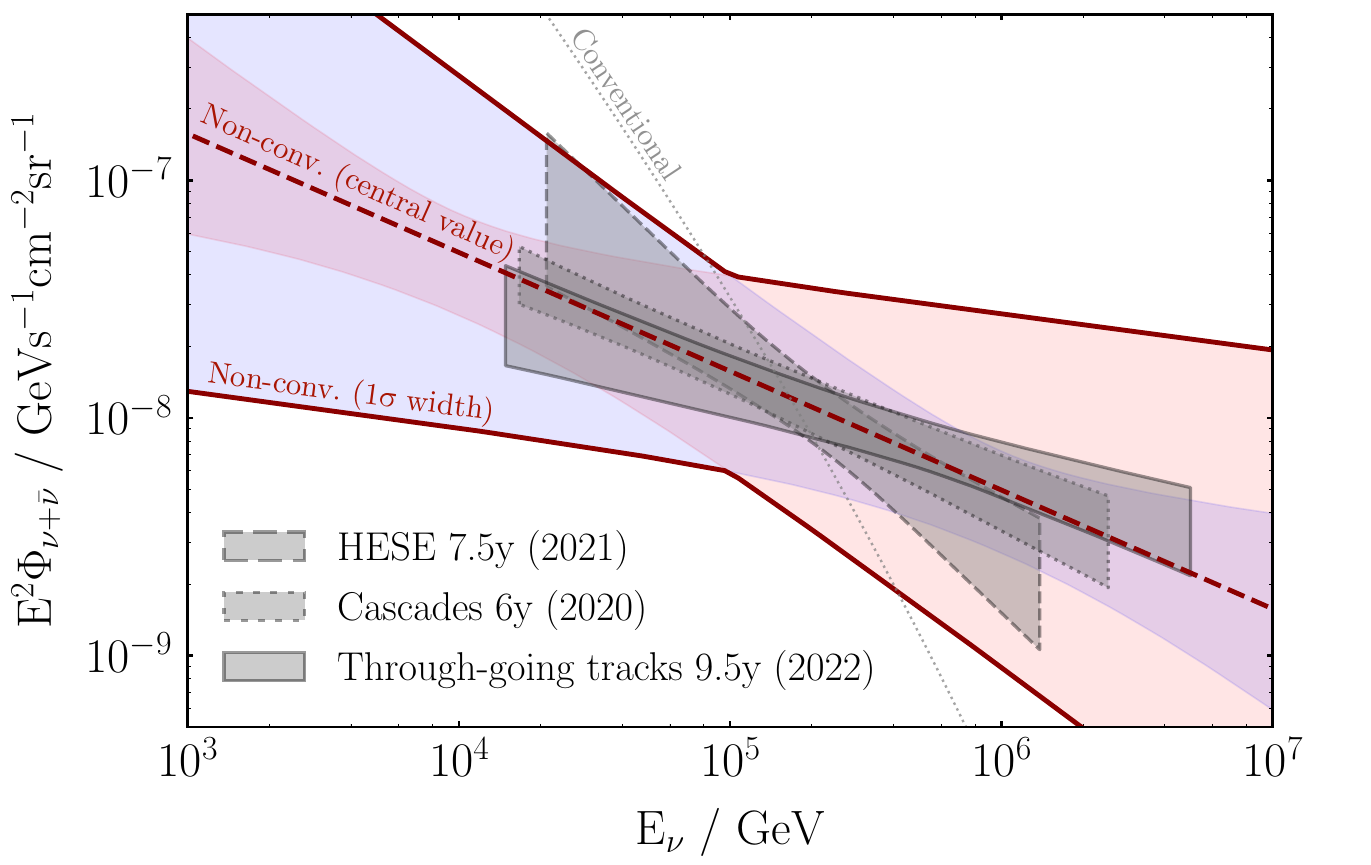}
\caption{ \textbf{Non-conventional flux systematic}. The shaded grey areas indicate the single-power-law energy spectra of the astrophysical neutrino flux measured by IceCube~\cite{IceCube:2020wum,IceCube:2021uhz, IceCube:2020acn}.
Red lines delineate the contour encompassed using the $1\sigma$  prior width from this analysis (blue and red shaded areas show the contours assuming two different values for the pivot energy parameter, 10~TeV or 1~PeV), with the dashed line denoting the nominal prediction. 
The dashed grey line represents the (sky-averaged) conventional $\nu_\mu+\bar{\nu}_\mu$ flux.}
\label{fig:nonconv}
\end{figure}

The non-conventional component encompasses both astrophysical and prompt neutrinos. 
The nominal prediction for the astrophysical component adopts a single power-law energy spectrum (with the same normalization and slope as in the previous analysis) and an isotropic angular distribution. 
For its flavor composition, we assume  $(\nu_\mu\!:\!\nu_e\!:\!\nu_\tau)=(1\!:\!1\!:\!1)$ and $(\nu\!:\!\bar{\nu})=(1\!:\!1)$. 
The prompt component is modeled using Sibyll2.3c~\cite{Riehn:2017mfm}, which is comparable to other contemporary perturbative QCD calculations~\cite{Fedynitch:2018cbl}. 
In this analysis, we expanded the energy range from 10 to 100~TeV, making the contributions from the possible prompt atmospheric and astrophysical neutrino fluxes more significant. 
Consequently, we adopted a more conservative approach to model the uncertainties associated with these components. 
Instead of a single power law we have employed a broken power-law with four parameters: a normalization term, two tilt terms, and an energy pivot point.
Figure~\ref{fig:nonconv} illustrates the energy spectrum region allowed with the prior widths of this analysis, designed to encompass IceCube's various astrophysical neutrino measurements.
Additionally, we conducted multiple studies to quantify the risk of spurious fits to sterile neutrino hypotheses arising from mismodeling of the non-conventional flux.
These tests included scenarios such as increasing the prompt component by an order of magnitude or removing it altogether, considering $\nu$- and $\bar{\nu}$-only astrophysical contributions, and incorporating a galactic component~\cite{IceCube:2023ame}. 
Our investigations revealed a significance for spurious signals below 0.3$\sigma$ in all test cases.

\subsection{Normalization}
\label{sec:norm}

In our previous analysis, a 40\% uncertainty was introduced in the normalization term for the conventional neutrino flux. 
Our updated treatment of this component now incorporates both shape and normalization uncertainties; therefore, we have revised the implementation of a normalization term.
Key changes include extending the normalization term to cover non-conventional components and adjusting the prior width to capture the influence of neutrino cross sections and muon energy losses in this analysis.

The neutrino cross section impacts interactions near the detector. 
In the TeV regime, the modeling of the neutrino-nucleon cross section relies on perturbative QCD. 
Our nominal prediction employs the CSMS model~\cite{Cooper-Sarkar:2011jtt}.
It has been noted that total cross-section predictions can differ by $\mathcal{O}(5\%)$ depending on the parton distribution functions used and target-dependent nuclear effects~\cite{Binder:2017rlx,Klein:2020nuk,Garcia:2020jwr,Xie:2023suk,Jeong:2023hwe,Candido:2023utz,Reno:2023sdm}. 
In light of this, we incorporated a 10\% uncertainty into the normalization budget.

Regarding muon energy losses, recent calculations indicate an overall 3\% uncertainty associated with dominant interaction channels in this energy regime, namely pair production, bremsstrahlung, and photo-hadronic processes~\cite{Sandrock_2020}.
The primary impact of this uncertainty lies in the muon range, affecting the assumed effective volume.
Therefore, to first order, this error would correspond to a 9\% uncertainty in the overall number of interactions triggering the detector.
Other effects, such as energy dependency and differences between starting and through-going events, are not currently considered.

Additional factors affecting the overall normalization include the bedrock density and the detector's livetime. However, uncertainties on these parameters are below 1\%.
Finally, we studied the impact of final state radiation~\cite{Plestid:2024bva} on the analysis, which alters the fraction of the energy from the neutrino carried out by the outgoing lepton.
We observed a significance for a spurious signal of 0.04$\sigma$ from the mismodeling of this effect.

In conclusion, the prior width of the normalization term has been updated to 20\%.

\subsection{Neutrino attenuation}
\label{sec:att}

At $E_{\nu}\sim40$\,TeV, the interaction length for neutrinos is roughly the diameter of the Earth.
The uncertainty associated with this phenomenon depends on our understanding of neutrino deep inelastic scattering and the Earth's density profile. 
Regarding the Earth's density profile, we use the Preliminary Reference Earth Model (PREM). Our investigation indicates that employing alternative models~\cite{10.1111/j.1365-246X.1995.tb03540.x,Moser1982THEIS} results in sub-percent changes in the reconstructed distributions of our analysis~\cite{Axani:2019sbk}. 

Concerning the uncertainty in cross section, we refer to Sec.~\ref{sec:norm} for a discussion on the prior width. However, neutrino interactions with heavy targets beyond 100~TeV primarily drive the attenuation effect, where shadowing effects become relevant.
Therefore, we assign nuisance parameters to model this effect for neutrinos and antineutrinos, treating them as uncorrelated with the normalization term.

\subsection{Bulk ice}
\label{sec:ice}

The presence of ice impurities, referred to as ``dust," between IceCube strings and the crystalline microstructure of the ice significantly influences the scattering and absorption of light. 
We address the uncertainty of dust concentration in different ice layers by leveraging LED flasher data~\cite{IceCube:2013llx}. 
To assess the impact of this effect, we conducted simulations using the SnowStorm method~\cite{IceCube:2019lxi}. 
This method characterizes variations in the ice model through correlated amplitudes and phases of Fourier modes.
The impact of these variations becomes negligible after the fourth mode in the reconstructed energy-zenith space. 

\begin{figure}[tb!]
    \centering
    \includegraphics[width=0.42\textwidth]{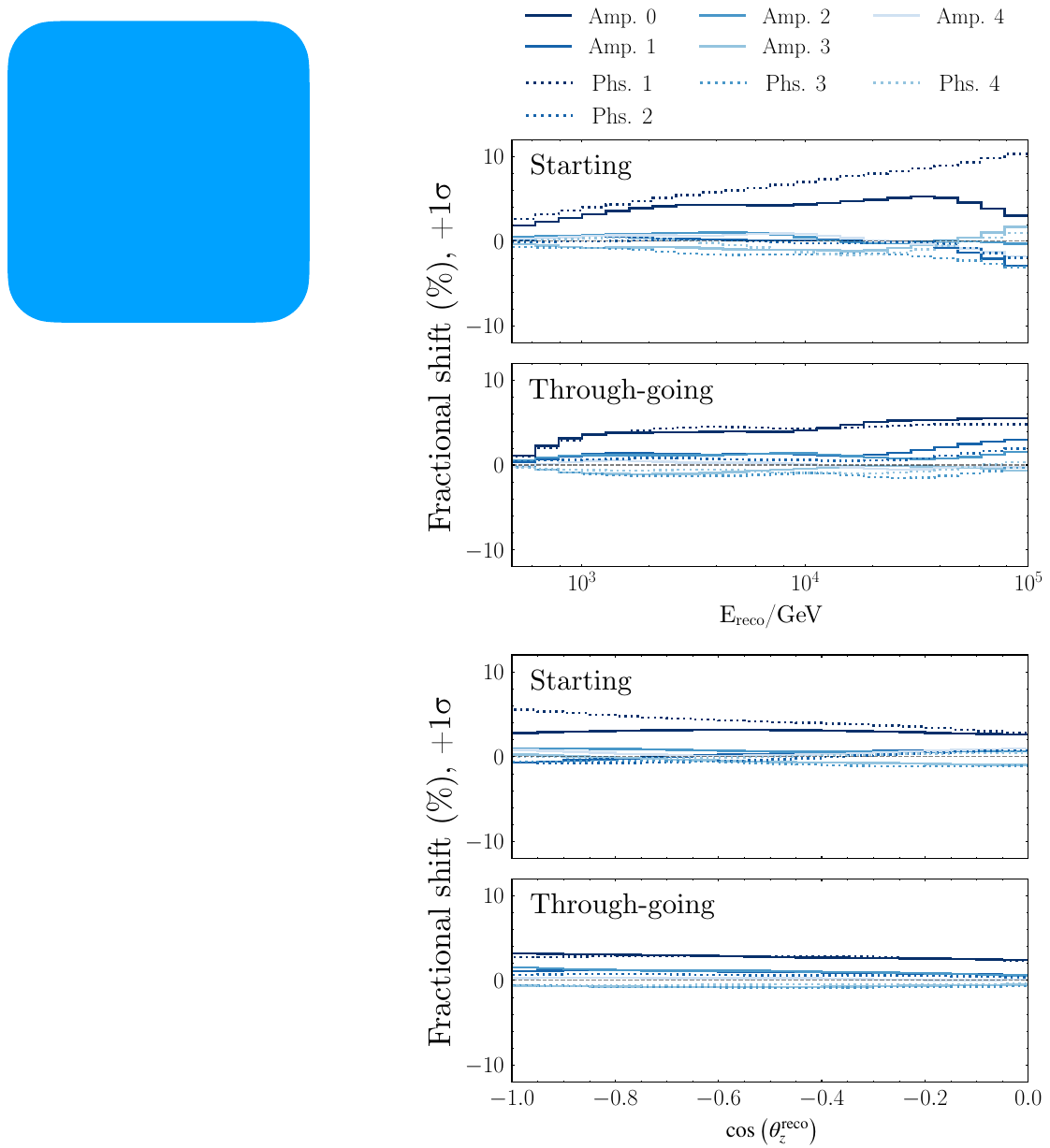}
    \caption{ \textbf{Bulk ice systematics}. The percentage-wise fractional differences in the expected number of events as a function of reconstructed energy (top panels) and zenith angle (bottom panels) when shifting by +1$\sigma$ the amplitude (solid; so-called \textit{Amp.}) and phases (dashed; so-called \textit{Phs.}) of Fourier modes describing variations in the bulk ice model. The color gradient shifts from darker to lighter blue as the mode increases.}
    \label{fig:icegrad}
\end{figure}

As illustrated in Fig.~\ref{fig:icegrad}, the primary impacts stem from the zeroth-mode amplitude and the first-mode phase (i.e., $\textit{Amp. 0}$ and $\textit{Phs. 1}$), while subsequent modes contribute second-order corrections. 
The $\textit{Amp. 0}$ sets the absolute scale of absorption and scattering in the detector, and the $\textit{Phs. 1}$ encodes the most relevant depth-dependent features.
The latter affects starting and through-going events differently. 
The main reason is that the light pattern is partially contained within the detector for starting events, making them more sensitive to depth-dependent ice properties.

\begin{figure}[tb!]
    \centering
    \includegraphics[width=0.4\textwidth]{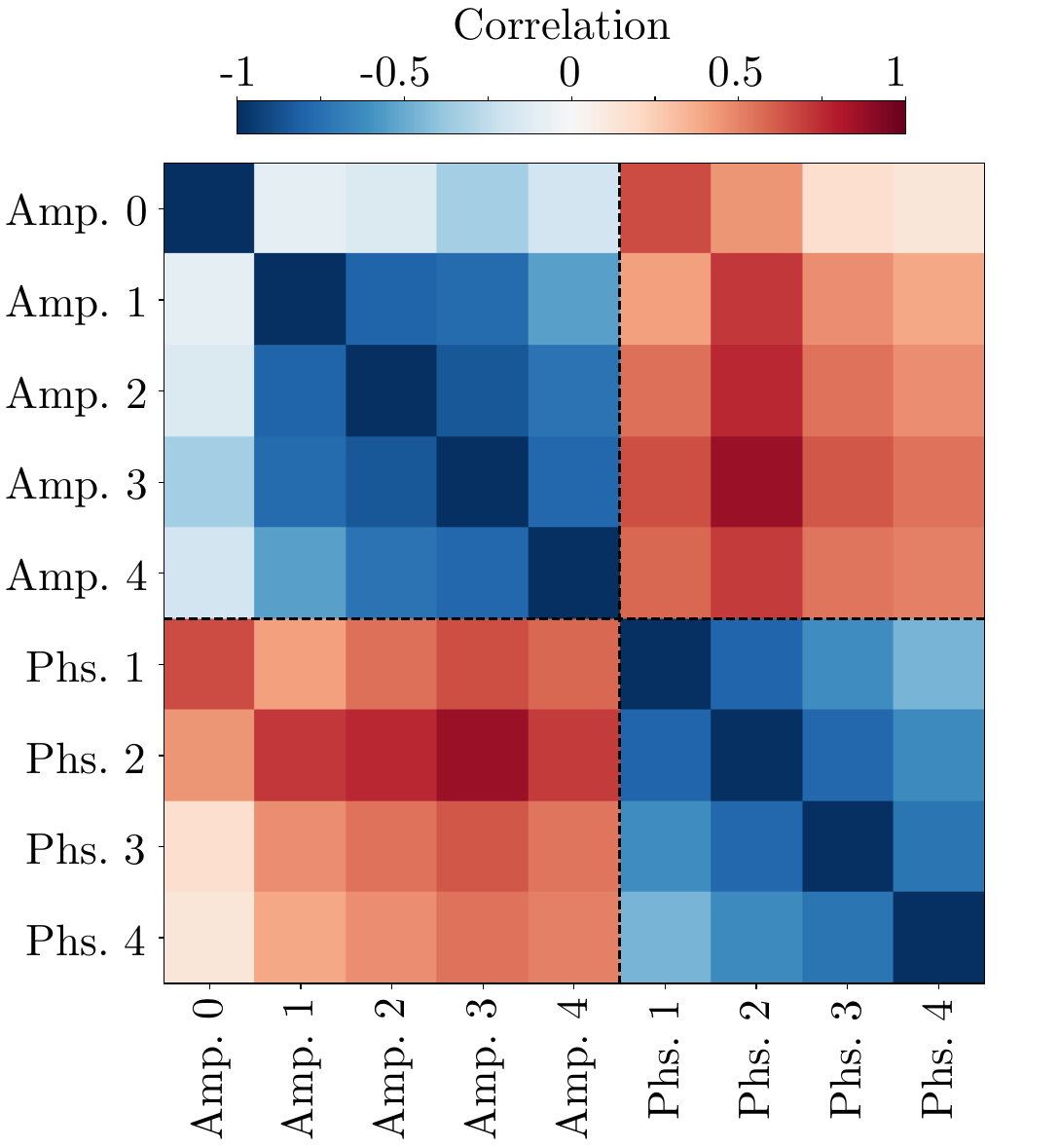}
    \caption{ \textbf{Correlation for bulk ice parameters}. Correlation matrix of the bulk ice parameters used in this analysis extracted from Ref.~\cite{IceCube:2019lxi}.}
    \label{fig:icecorr}
\end{figure}

Consequently, our analysis incorporates five amplitude and four phase parameters as nuisance parameters, including a penalty term with correlated priors as shown in Fig.~\ref{fig:icecorr}. Additionally, we found that using simulation produced with the birefringent ice properties incorporated in the bulk ice model~\cite{IceCube:2024qxf} has negligible impact on this analysis.

\subsection{Local response of the DOMs}
\label{sec:local}

The photon detection efficiency of the DOMs and the characteristics of the refrozen ice in the boreholes, or ``hole ice," are parameterized using the same methodology as in our previous analysis~\cite{IceCube:2020tka}. 
We conducted a series of simulations, introducing variations in both refrozen ice properties and DOM efficiencies. 
The resulting fractional differences in the reconstructed energy-zenith space were computed employing splines, differentiating between starting and through-going events. 
To account for uncertainties in these parameters, a nuisance parameter is assigned to each, with wide priors consistent with those employed in our earlier analysis. The allowed range for each parameter in the fit is defined by its minimum and maximum values assumed in the simulations to prevent extrapolations.

\subsection{Effect on sensitivity}

These uncertainties are incorporated into the likelihood through nuisance parameters with penalty terms.
The parameters describing our model for each are listed in Table~\ref{tab:systematics} by category.
The ``central value'' corresponds to the nominal value of the parameters in the simulations described above.
The uncertainties are assumed to be Gaussian with the 1$\sigma$ width quoted in column 3 of the table (with the exception of the pivot energy of the non-conventional flux).
The allowed range for the fit is indicated in column 4.

\begin{figure}[t!]
    \centering
    \includegraphics[width=0.5\textwidth]{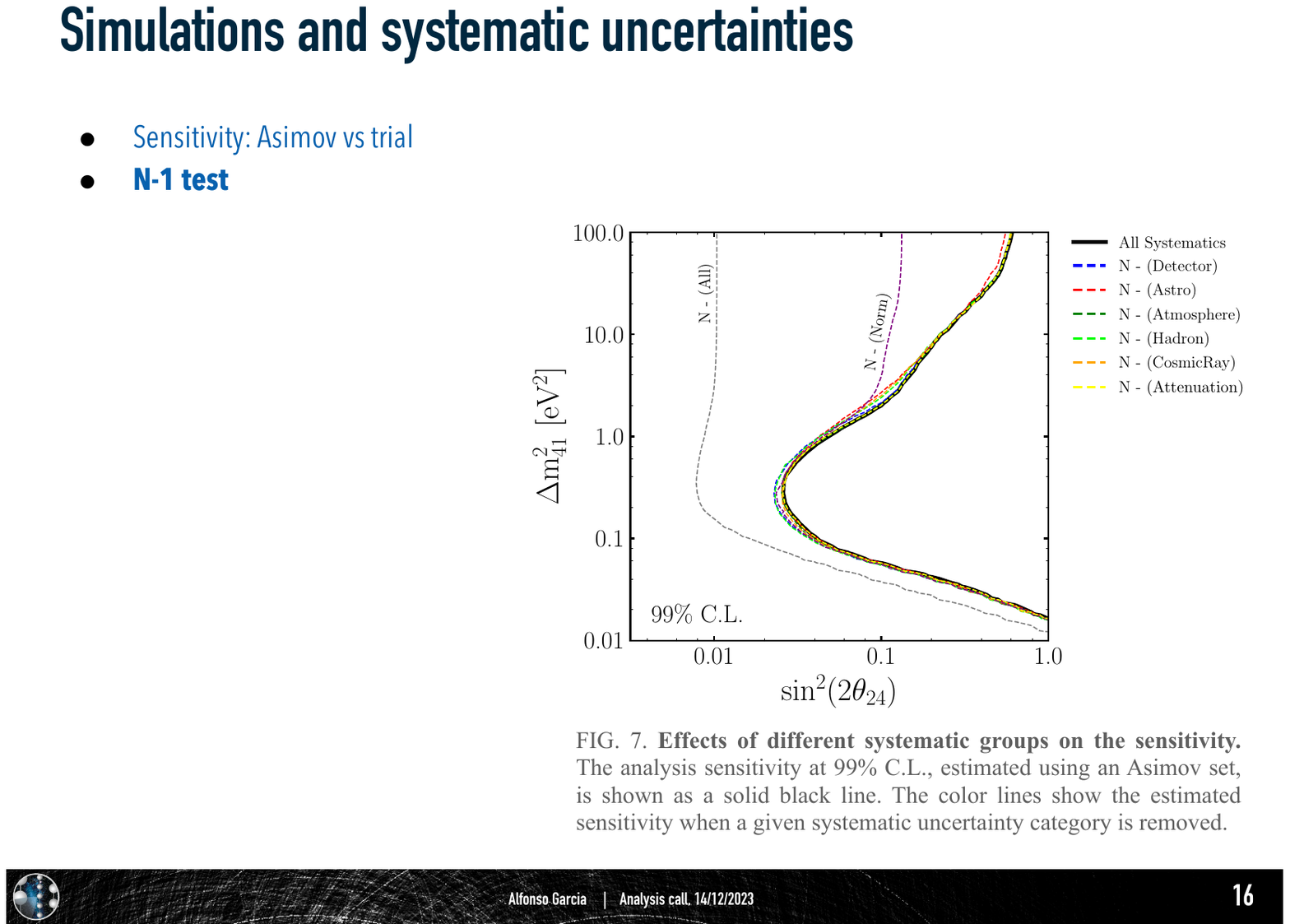}
    \caption{ \textbf{Effects of different systematic groups on the sensitivity}. The analysis sensitivity at 99\% C.L., estimated using an Asimov set, is shown as a solid black line. The dashed lines show the estimated sensitivity when a given systematic uncertainty category is removed.}
    \label{fig:n1_test}
\end{figure}

\begin{table*}[b!]
\footnotesize
\centering
\begin{tabular}{ll|r|r|r||r|r|r}
& Nuisance &  Central & 1$\sigma$ width & ~~~~~Allowed & Pull Null & Pull Best & Pull Difference\\
& parameter & value & of prior & range & Fit ($\sigma$) &  Fit ($\sigma$) & $|$Null-Best Fit$|$ ($\sigma$)  \\ 
\hline
\multicolumn{8}{l}{Overall normalization (Sec.~\ref{sec:norm})} \\
\hline
& Norm & 1.00& 0.2 & 0.10,3.00 & -0.05 & 0.41 & 0.46\\
\hline
\multicolumn{8}{l}{Local response of DOMs (Sec.~\ref{sec:local})} \\
\hline
& DOM efficiency & 1.00& 0.10 & 0.97,1.06 & 0.02 & 0.03 & 0.01\\
& Forward hole ice & -1.00& 10.00 & -5.35,1.85 & 0.28 & 0.27 & 0.01\\
\hline
\multicolumn{8}{l}{Bulk ice (Sec.~\ref{sec:ice})} \\
\hline
& Amplitude 0 & 0.00 & 1.00&  -3.00,3.00& 0.64 & 0.69 & 0.05\\ 
& Amplitude 1 & 0.00 & 1.00&  -3.00,3.00& 1.36 & 1.19 & 0.17\\
& Amplitude 2 & 0.00 & 1.00&  -3.00,3.00& 1.35 & 1.42 & 0.07\\
& Amplitude 3 & 0.00 & 1.00&  -3.00,3.00& 0.74 & 0.75 & 0.01\\
& Amplitude 4 & 0.00 & 1.00&  -3.00,3.00& 1.12 & 1.16 & 0.04\\
& Phase 1 & 0.00 & 1.00&  -3.00,3.00& -1.60 & -1.67 & 0.07\\
& Phase 2 & 0.00 & 1.00&  -3.00,3.00& -0.59 & -0.54 & 0.05\\
& Phase 3 & 0.00 & 1.00&  -3.00,3.00& -0.21 & -0.08 & 0.13\\
& Phase 4 & 0.00 & 1.00&  -3.00,3.00& 0.10 & 0.27 & 0.17\\
\hline
\multicolumn{8}{l}{Conventional flux (Sec.~\ref{sec:conv})} \\
\hline
& Atm. density ($\rho_{\textrm{atm}}$) & 0.00 & 1.00& -3.00,3.00 & -0.48 & -0.55 & 0.07 \\
& Kaon energy loss ($\sigma_{\textrm{K-Air}}$) & 0.00 & 1.00& -3.00,3.00 & 0.66 & 0.51  & 0.15\\
\multirow{6}{*}{\rotatebox[origin=c]{90}{Hadronic production$|$}} & K$_{158G}^{+}$ & 0.00 & 1.00& -2.00,2.00 & 0.93 & 0.89 & 0.04\\
 & K$_{158G}^{-}$ & 0.00 & 1.00& -2.00,2.00 & 0.29 & 0.24 & 0.05\\
 & $\pi_{20T}^{+}$ & 0.00 & 1.00& -2.00,2.00 & 0.15 & -0.06 & 0.21\\
 & $\pi_{20T}^{-}$ & 0.00 & 1.00& -2.00,2.00 & 0.17 & -0.03 & 0.20\\
 & K$_{2P}^{+}$ & 0.00 & 1.00& -2.00,2.00 & 0.28 & 0.09 & 0.19\\
 & K$_{2P}^{-}$ & 0.00 & 1.00& -1.50,2.00 & 0.24 & 0.01 & 0.23\\
 & $\pi_{2P}^{+}$ & 0.00 & 1.00& -2.00,2.00 & -1.50 & -1.23 & 0.27\\
 & $\pi_{2P}^{-}$ & 0.00 & 1.00& -2.00,2.00 & -1.08 & -0.85 & 0.23\\
 & p$_{2P}$ & 0.00 & 1.00& -2.00,2.00 & -0.25 & -0.18 & 0.07\\
 & n$_{2P}$ & 0.00 & 1.00& -2.00,2.00 & -0.17 & -0.15 & 0.02\\
\multirow{6}{*}{\rotatebox[origin=c]{90}{CR spectrum$|$}} & GSF$_{1}$ & 0.00 & 1.00& -4.00,4.00 & -0.33 & 0.10 & 0.43\\
 & GSF$_{2}$ & 0.00 & 1.00& -4.00,4.00 & -0.12 & -0.28 & 0.16\\
 & GSF$_{3}$ & 0.00 & 1.00& -4.00,4.00 & -0.12 & -0.05 & 0.07\\
 & GSF$_{4}$ & 0.00 & 1.00& -4.00,4.00 & -0.13 & -0.25 & 0.12\\
 & GSF$_{5}$ & 0.00 & 1.00& -4.00,4.00 & 1.82 & 2.24 & 0.42\\
 & GSF$_{6}$ & 0.00 & 1.00& -4.00,4.00 & -1.17 & -1.31 & 0.14\\
\hline
\multicolumn{8}{l}{Non-conventional flux (Sec.~\ref{sec:nonconv})} \\
\hline
& $\Phi^{\mathrm{HE}}/10^{-18}\mathrm{GeV}^{-1}\mathrm{sr}^{-1}\mathrm{s}^{-1}\mathrm{cm}^{-2}$ & 0.787 & 0.36 & 0.00,3.00 & 0.25 & 0.61 & 0.36\\
& $\log_{10}$ of pivot energy, $E_{\mathrm{break}}^{\mathrm{HE}}/\mathrm{GeV}$  & - & - & 4.00,6.00 & *4.25 & *4.31  & N/A, see caption\\
& $\Delta\gamma_{1}^{\mathrm{HE}}$, tilt from -2.5 & 0.00 & 0.36 & -2.00,2.00 & 2.62 & 2.39 & 0.23\\
& $\Delta\gamma_{2}^{\mathrm{HE}}$, tilt from -2.5 & 0.00 & 0.36 & -2.00,2.00 & -0.22 & 0.10 & 0.21\\
\hline
\multicolumn{8}{l}{Neutrino attenuation (Sec.~\ref{sec:att})} \\
\hline
& $\nu$ attenuation & 1.00& 0.10 & 0.82, 1.18 & 0.12 & -0.14 & 0.26\\
& $\bar{\nu}$ attenuation & 1.00& 0.10 & 0.82, 1.18 & 0.04 & -0.02 & 0.06\\ \hline
\end{tabular}
\caption{List of systematic parameters considered in this analysis along with their priors and allowed ranges.   The pull results for the null and best fits, measured in $\sigma$, with sign, are also listed for comparison (except for $E_{\mathrm{break}}^{\mathrm{HE}}$ for which the fitted value, indicated with $*$, is provided).}
\label{tab:systematics}
\end{table*}

Fig.~\ref{fig:n1_test} quantifies the influence of each uncertainty source on the expected Asimov sensitivity for this analysis.
Each curve represents a scenario where the nuisance parameters linked to one type of uncertainty are held constant at their central values while allowing others to fluctuate within their constraints. 
At large $\Delta m^2_{41}$, the most significant impact stems from normalization, followed by astrophysical and cosmic-ray flux uncertainties. 
Conversely, at low $\Delta m^2_{41}$, the dominant factors are uncertainties in the detector and hadronic yields.
While removing most of the systematic uncertainties individually exhibits minimal impact on sensitivity, the collective effect of systematics is more substantial when all are removed simultaneously.

\section{Fitting the 3+1 model}

The 3+1 model in IceCube produces two distinct and related signatures.
The first is vacuum oscillations appearing in the neutrino and antineutrino fluxes.
The second is the matter resonance, which produces a deficit in only the antineutrino flux.
Both of these oscillation features are numerically calculated consistently by means of the nuSQuIDS package~\cite{Arguelles:2021twb}.
In what follows we discuss the these two relevant features.

This analysis assumes that $\theta_{14}=\theta_{34}=0$ and leaves $\theta_{24}$ and $\Delta m^2_{41}$ as parameters in the fit.
Fitting for more than two parameters is computationally costly, and so we begin this iteration of studies with only these two, recognizing that this is, in fact, a simplified 3+1 model.
The impact of $\theta_{14}$ is negligible in this analysis since the $\nu_e/\nu_\mu$ ratio is higher than 20 in the TeV regime~\cite{Yanez:2023lsy}, whereas setting $\theta_{34}=0$ yields conservative constraints on $\theta_{24}$~\cite{Esmaili:2013vza}.

With this approximation, the vacuum oscillation signature is given by 
\begin{equation} 
P_{\nu_\mu \rightarrow \nu_\mu} = \sin^2(2\theta_{24})\sin^2 (1.27 \Delta m_{41}^2 L/E) \\
\end{equation}

\begin{figure}[tb!]
    \centering
    \includegraphics[height=0.5\textheight]{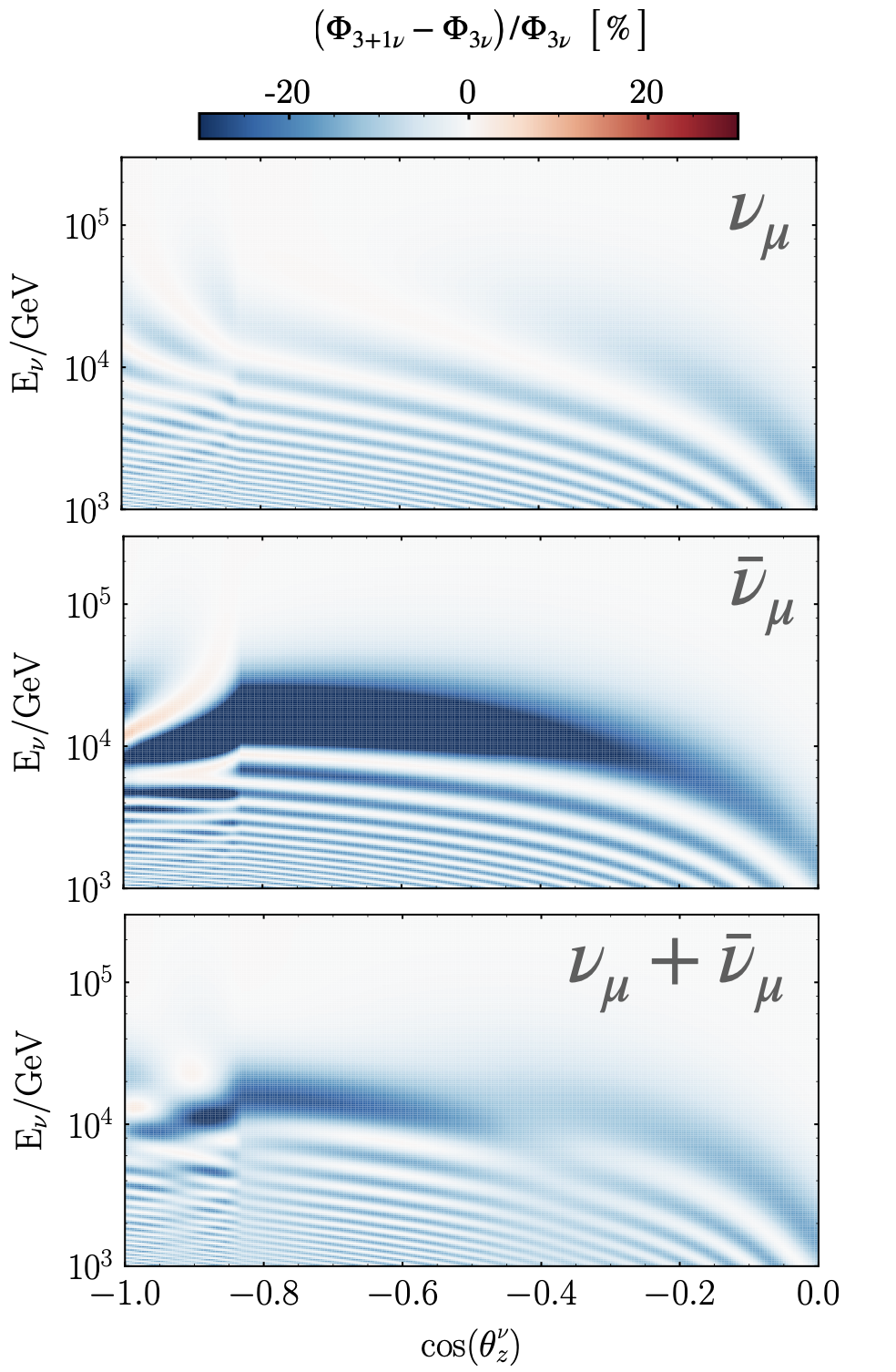}
    \caption{Fractional differences of the atmospheric flux predictions between the best-fit and null hypotheses.
    The plots use the true neutrino energy and direction. The best-fit values used and $\sin^2(2\theta_{24}) = 0.16$ and $\Delta m^{2}_{41} = 3.5$~eV$^2$.}
    \label{fig:oscillresult}
\end{figure}

Because the zenith angle depends upon $L$, the oscillation deficit will trace arcs in the $E_\nu$ versus $\cos(\theta_z^{\nu})$ space, as seen in the ``oscillogram'' in Fig.~\ref{fig:oscillresult}, which shows the ratio of 3+1 $\nu_\mu$ disappearance to the null hypothesis for the best-fit parameters of this analysis.
The deficit corresponds to the first oscillation maximum arcs across the top right.

The matter resonance arises when one includes a matter potential in the Hamiltonian describing oscillations.
This modifies the oscillation amplitude such that it becomes very large for
\begin{equation}
    E_\nu^{\mathrm{resonance}} = \mp (\Delta m^2_{41} \cos(2 \theta_{24})/(\sqrt{2}G_F N_n),
\end{equation}

\noindent where $G_F$ is the Fermi constant and $N_n$ is the neutron number density.
The sign is negative for neutrinos and positive for antineutrinos.
If $\theta_{24}<\pi/4$ and $\Delta m_{41}^2$ is positive, which is required to respect bounds from many experiments, then the resonance will appear in the antineutrino flux.
As a result, the antineutrino events will exhibit both vacuum oscillations and a resonance, as seen in Fig.~\ref{fig:oscillresult} (middle).    

In this analysis, the selected events encompass neutrino and antineutrino interactions.
Therefore, the effects add according to the relative contribution of $\nu_\mu$ and $\bar \nu_\mu$, which, in the range of the conventional flux, is approximately 10 to 1.
Fig.~\ref{fig:oscillresult} bottom shows the properly summed event rates.
One sees that the resonance adjoins the first oscillation maximum, leading to a significant deficit in the form of an arc spanning $\cos(\theta_z^{\nu})$.

\begin{figure}[tb!]
    \centering
    \includegraphics[width=0.5\textwidth]{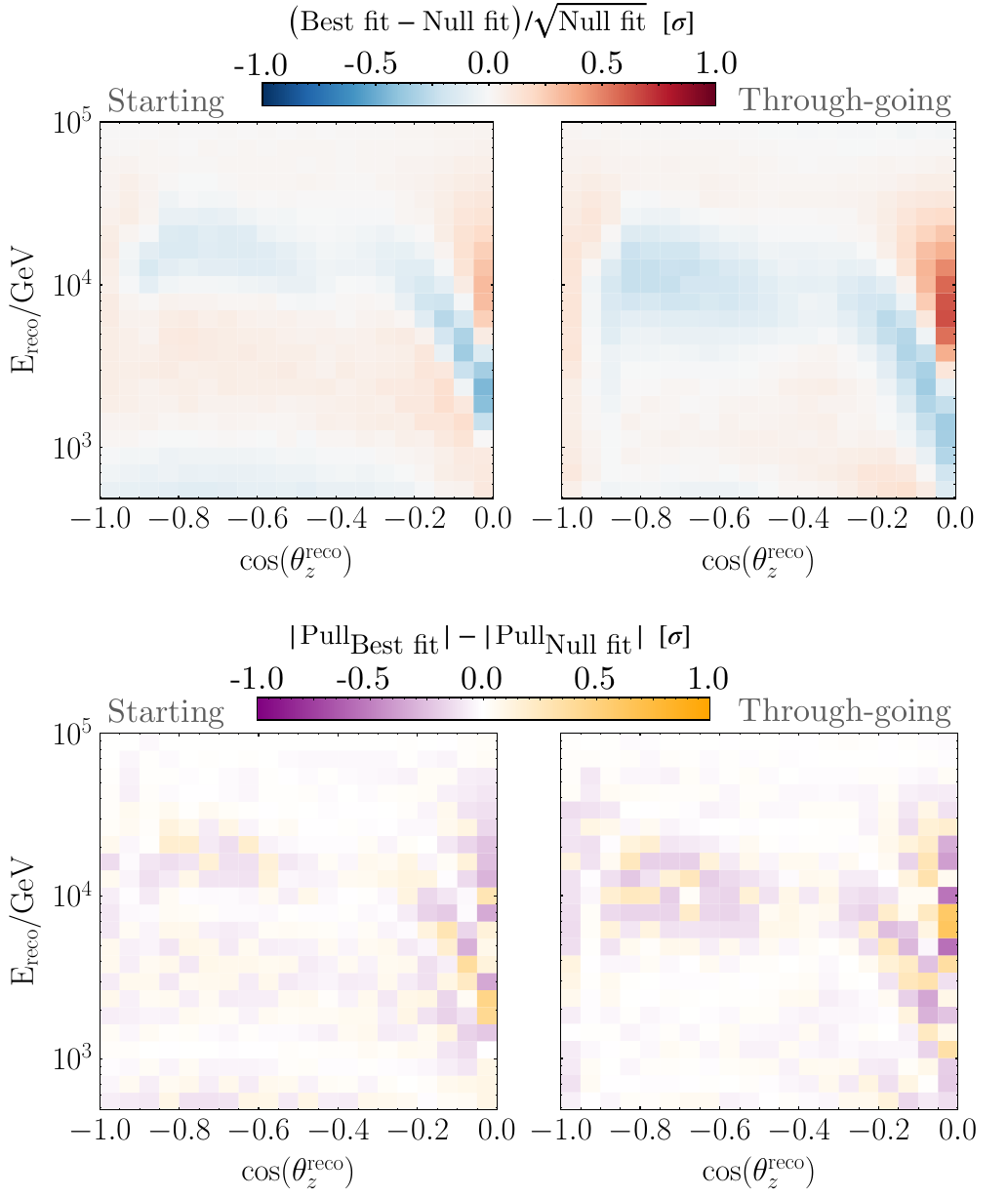}
    \caption{\textbf{Expected and Observed Signal}. 
    \textit{Top panels}: Comparison of the best-fit and null hypothesis expectations for reconstructed starting and through-going events. 
    Red (blue) colors indicate an excess (deficit) of events in the best-fit prediction relative to the null hypothesis. 
    \textit{Bottom panels}: Difference between data pulls for the best-fit values ($\sin^2(2\theta_{24}) = 0.16$ and $\Delta m^{2}_{41} = 3.5$~eV$^2$) and null hypotheses for the starting and through-going samples. 
    Purple indicates the best-fit is preferred in a given bin; orange indicates a preference for the null hypothesis.}
    \label{fig:cricles}
\end{figure}

Resolution on $E_\nu$ and $\cos(\theta_z^{\nu})$ will smear the signature seen in the bottom oscillogram of Fig.~\ref{fig:oscillresult}.
To see the effect, we compare the predictions for the best-fit and null models using the fitted nuisance parameters in each case.
The bin-by-bin pull of the best fit minus the null prediction divided by the square root of the null prediction produces the prediction shown in the top panels of Fig.~\ref{fig:cricles} for starting and through-going events.  
The apparent excess occurs because the normalization is allowed to float in the fit.
As indicated in Table~\ref{tab:systematics}, the normalization for the best-fit hypothesis is higher than that for the null hypothesis.
This difference arises because the fast oscillations reduce the expected flux below the resonance, as illustrated in Fig.\ref{fig:oscillresult}.
In addition, one can see that the arc-like feature of the combined resonance and first oscillation maximum is retained even after detector effects are included.
Detector smearing leads to an extended no-oscillation region below the arc.
From 0.5 to 1~TeV, particularly in the case of starting events, an overall deficit from the fast oscillations can be seen at $\cos(\theta_z^{\mathrm{reco}})<-0.2$.

\subsection{Frequentist Fit Results \label{sec:freq}}

The frequentist result was obtained following a blind analysis. 
The best-fit point was found at $\sin^2(2\theta_{24}) = 0.16$ and $\Delta m^{2}_{41} = 3.5$~eV$^2$.
Compared to the no sterile neutrino hypothesis, the test statistic was $-2\Delta \log \mathcal{L}=6.96$, corresponding to a p-value of 3.1\% for two degrees of freedom. 

\begin{figure*}[h!]
    \centering
    \includegraphics[width=0.9\textwidth]{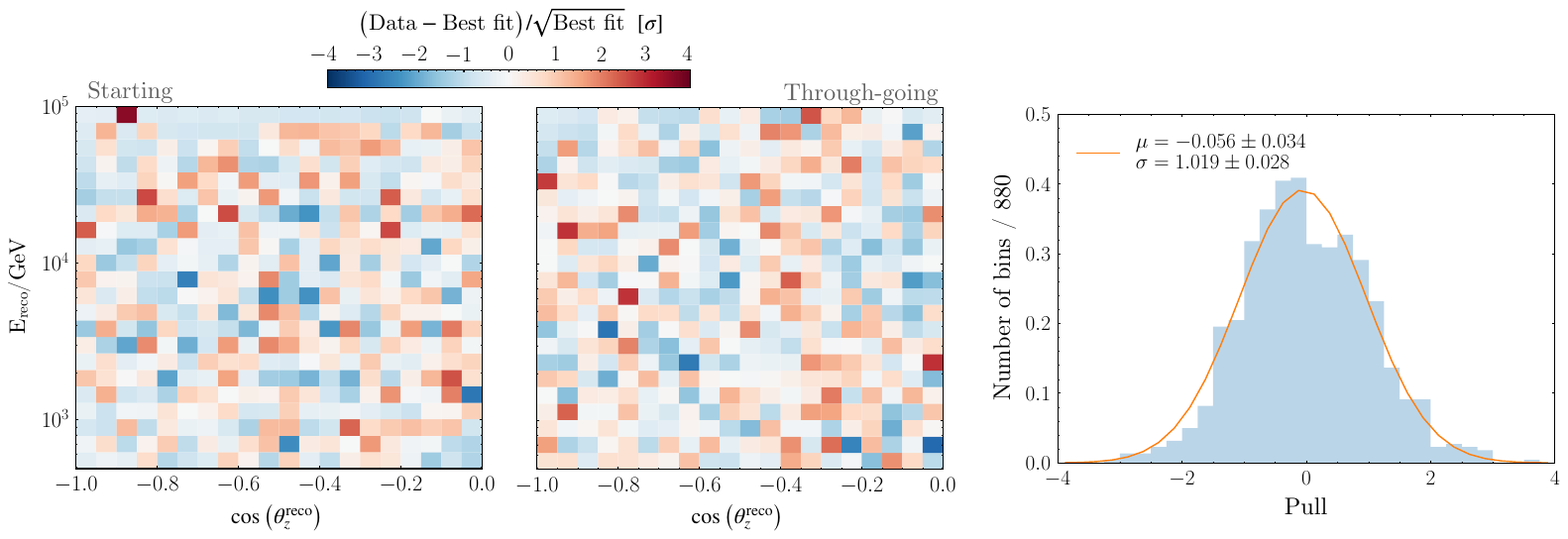}
    \caption{\textbf{Two- and one-dimensional pulls}. \textit{Left panels}: pulls between data and best-fit expectations as a function of energy and zenith. \textit{Right panel}: One-dimensional distribution of the pulls from the 880 bins used in this analysis. The orange line represents the Gaussian fit.}
    \label{fig:2d_pulls}
\end{figure*}

\begin{figure*}[h!]
    \centering
    \includegraphics[width=1.0\textwidth]{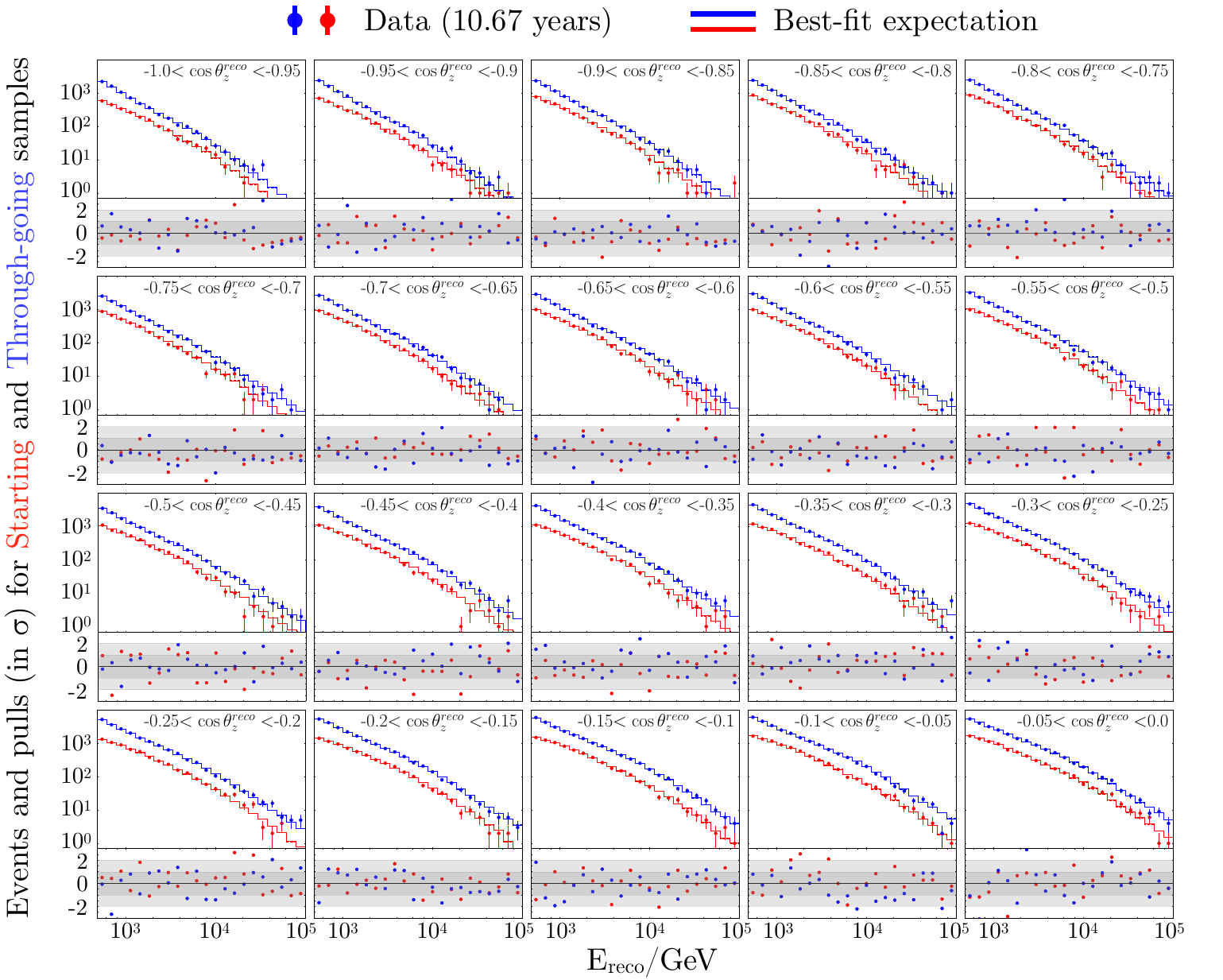}
    \caption{\textbf{One-dimensional distributions}. Energy distribution for each zenith for starting (red) and through-going events (blue). Data points are shown as markers, with bars representing the statistical error. The solid lines show the prediction from the best-fit sterile neutrino hypothesis. The bottom panels show the statistical pulls between the observed and expected distribution, with grey bands indicating the $\pm1$ and $\pm2\sigma$ bands.}
    \label{fig:1d}
\end{figure*}

To further check for systematic disagreements between the prediction and the data,
Fig.~\ref{fig:2d_pulls} displays the data pull relative to the expectation under the best-fit hypothesis in each analysis bin.
These pulls show a random distribution, as seen visually bin-by-bin in Fig.~\ref{fig:2d_pulls}.
While several bins at the highest energy in the starting sample show no events (see Fig.~\ref{fig:data_2d}), this observation is consistent with expectations.
The distribution of bin-wise pulls was fitted to a Gaussian, which aligns well with a normal distribution ($\mu=-0.06\pm0.03$ and $\sigma=1.02\pm0.03$).
The agreement between data and expectation can also be visualized in the one-dimensional histograms for the reconstructed energy in different zenith bins shown in Fig.~\ref{fig:1d}.

The bottom panels in Fig.~\ref{fig:cricles} compare the bin-wise pulls between the data and both the best-fit and null expectations.
Purple (orange) bins indicate better agreement with the best-fit prediction (null hypothesis).
Across all bins, differences remain within the $\pm1\sigma$ range, with the largest deviations of $\pm0.6\sigma$, primarily observed in the most horizontal events within the through-going sample---a point discussed further in Sec.~\ref{sec:splits}.
Up to statistical fluctuations, the areas showing better agreement with the best-fit prediction coincide with regions where the pulls between the best-fit and null hypothesis are more prominent, as shown in the top panels in Fig.~\ref{fig:cricles}.
The main feature of the data that leads the fit to prefer the best-fit over the null hypothesis appears in the through-going sample for $\cos(\theta_z^{\textrm{reco}})<-0.4$ at energies around 10\,TeV. 
In fact, we observe that the null rejection increases when studying these regions separately, as discussed in Sec.~\ref{sec:splits}.

The pulls for the nuisance parameters associated with this frequentist study are quantified for the null and best fit in Table~\ref{tab:systematics} in the rightmost columns.
The comparison shows that the pulls are nearly identical in the two cases.
Thus, there is relatively little correlation between the systematic effects and the 3+1 model parameters. 

\begin{figure}[b!]
    \centering
    \includegraphics[width=0.5\textwidth]{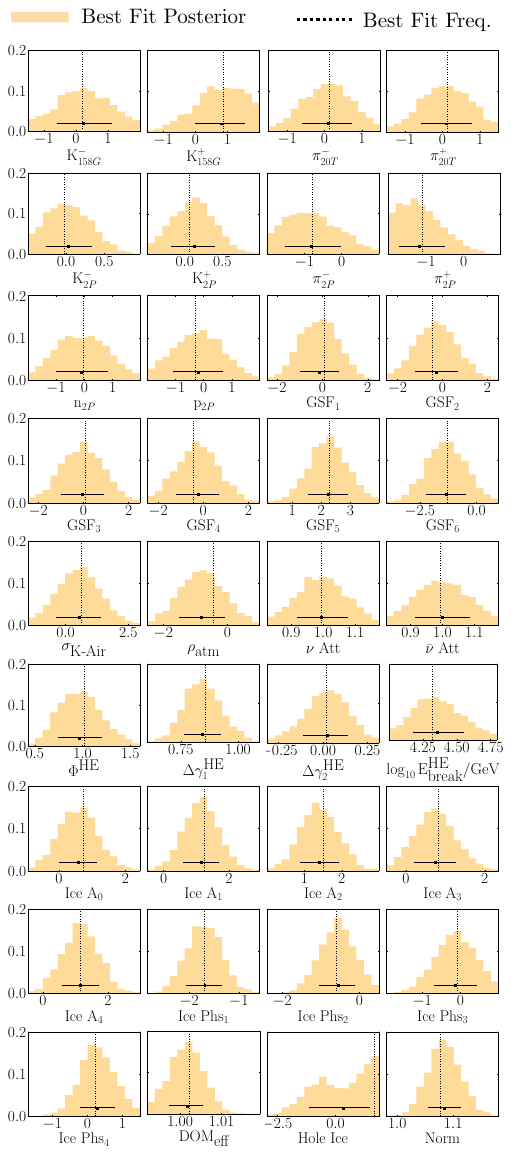}
    \caption{\textbf{Systematic posteriors}. Bayesian posterior distributions for the best-fit hypothesis. Vertical lines show the fitted value in frequentist analysis. Horizontal lines show the posteriors' mean and 1$\sigma$ width values.}
    \label{fig:posteriors}
\end{figure}

\subsection{Bayesean Fit Results}

To complement the frequenstist result, we have also developed a Bayesian analysis.
We have followed the same method as in the previous analysis~\cite{IceCube:2020tka}, which avoids dependence on the physics parameter priors as described in Ref.~\cite{Gariazzo:2019xhx}.
In this approach, the model evidence is computed for each sterile neutrino hypothesis by integrating the likelihood over the nuisance parameters using MultiNest~\cite{Feroz:2008xx}.
Then, we compare each sterile point in our ($\sin^2(2\theta_{24}),\Delta m^{2}_{41}$) grid to the null hypothesis.
We use the same priors as in the frequentist analysis for the nuisance parameters.

\begin{figure}[b!]
    \centering
    \includegraphics[width=0.46\textwidth]{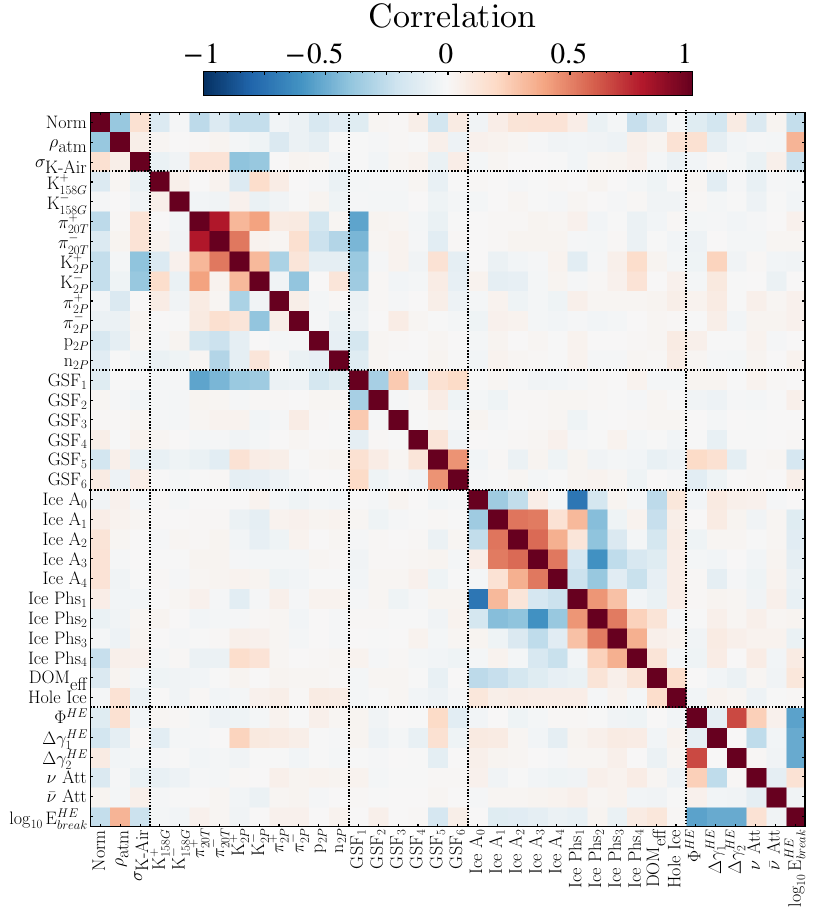}
    \caption{\textbf{Correlations}. Bayesian posterior correlation for systematics assuming the best-fit hypothesis. Dashed lines group systematics uncertainties associated with hadronic yields, cosmic-ray spectrum, detector effects, and non-conventional flux.}
    \label{fig:syst_corr}
\end{figure}

The nuisance parameters are fitted to data at each point in the parameter space. 
Fig.~\ref{fig:posteriors} shows each nuisance parameter's posterior distribution and fitted value at the best-fit point for the Bayesian and frequentist analyses. 
Fig.~\ref{fig:syst_corr} illustrates the correlations between nuisance parameters at the best-fit point.
Noteworthy correlations emerge among subsets of nuisance parameters, all falling within expected ranges. 
For instance, correlations between nuisance parameters with correlation priors, such as hadronic yields, cosmic ray, and bulk ice properties, are anticipated and observed.
Additionally, correlations are observed between DOM efficiency and bulk ice parameters.
Non-conventional flux parameters also exhibit correlations.

\begin{figure}[t!]
    \centering
    \includegraphics[width=0.5\textwidth]{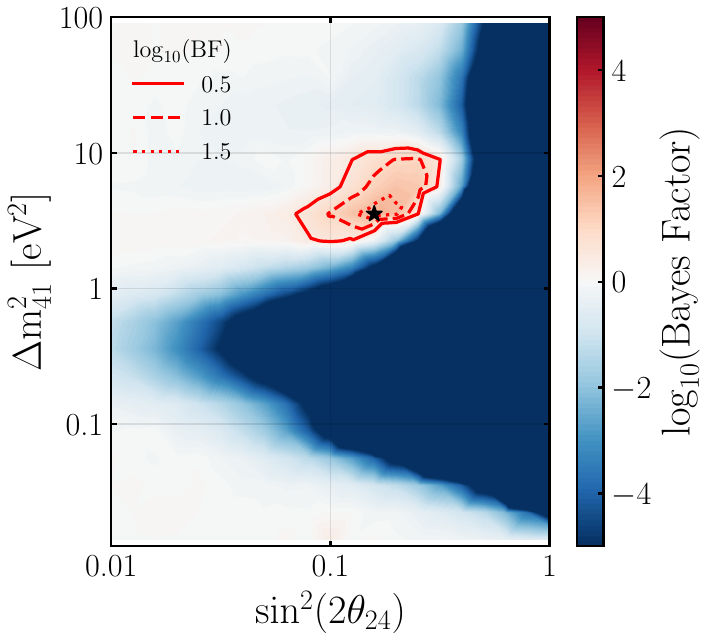}
    \caption{\textbf{Bayesian result}. The color scale shows the log10 of the BF relative to the null hypothesis. 
    The dotted, dashed, and solid black contours correspond to $\log_{10}(\textrm{BF})$ values of $0.5$, $1.0$, and $1.5$. 
    The best model is marked with a black star. 
    Systematic uncertainties are marginalized for each point $(\sin^2(2\theta_{24}),\Delta m^{2}_{41})$ in the scan.}
    \label{fig:bayes}
\end{figure}

Fig.~\ref{fig:bayes} shows the Bayes factor (BF) maps, where the point-wise BF is calculated relative to the no-sterile neutrino hypothesis. The best-model location, with a BF of 51.6, agrees with the best-fit point in the frequentist analysis. Contours are drawn in logarithmic Bayes Factor steps of 0.5, quantifying the strength of evidence~\cite{Jeffreys:1939xee}.

\newpage
\section{Post-unblinding stability tests \label{sec:splits}}

Once the results were unblinded, we performed multiple studies to investigate the stability of the result. 

\begin{figure*}[tb!]
    \centering
    \includegraphics[width=1.0\textwidth]{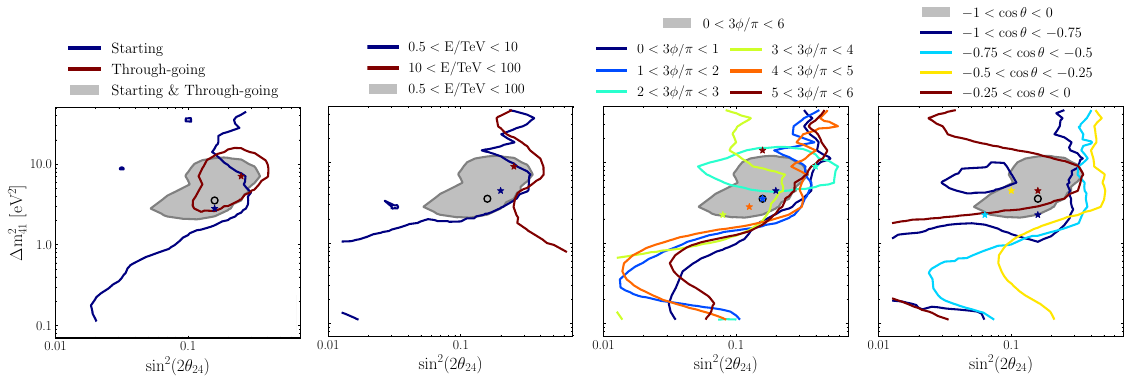}
    \caption{\textbf{Split sample fits}.
    Each plot corresponds to a reconstructed variable that has been divided into different regions. 
    Each color line shows the 95\% C.L. in a given range, and the stars represent the best-fit point. 
    The grey region shows the final result of this analysis, and the black circle is the best-fit point.
    \textit{From left to right}: starting vs. through-going, low vs. high energy, six azimuthal angles, and four zenith angles (note: the island observed in the contour for $-1<\cos\theta<-0.75$ indicates exclusion at 95\% C.L.).
    }
    \label{fig:threesplits}
\end{figure*}

By splitting the sample using various criteria and performing the fit separately on each, we can identify if there are regions of the reconstructed phase space where null hypothesis rejection is stronger or weaker and test the reproducibility of the best-fit point.
Hence, this is a valuable test to assess the consistency of the result, given our current understanding of the detector and the physical processes involved in this analysis.
We will compare to the 95\% C.L. allowed region for the full data set, which is shown in gray, with a gray star for the best-fit point, in Fig.~\ref{fig:threesplits}. 
%In addition, we will show the best-fit value for the nuisance parameters in each split. 

When interpreting these results, it is important to consider two factors. 
Firstly, sample splitting reduces statistical power in each dataset, resulting in 95\% C.L. limits rather than allowed regions in many cases. 
Secondly, the 95\% C.L. allowed region for the entire sample exhibits a flat likelihood, meaning that best-fit points for split samples are expected to move within or near this region.

Figure~\ref{fig:threesplits} (first panel) compares the results if starting and through-going events are fit separately compared to the simultaneous fit used in the final result.
The larger through-going event sample (red), with 274,309 events, retains a 95\% C.L. contour that overlaps the final fit well and has a best fit contained within the final sample 95\% C.L. allowed region.
The smaller starting sample (blue), with 93,762 events, does not have a 95\% C.L. closed contour. However, the best fit agrees with the final allowed region.
Also, the 95\% C.L. limit contour follows the final best-fit region, as expected.
%The pulls of the nuisance parameters agree within 2$\sigma$ except for the zeroth-mode amplitude of the bulk ice.
%Finally, the goodness-of-fit p-value is 31\% and 37\% for starting and through-going samples, respectively.

Figure~\ref{fig:threesplits} (second panel) divides the sample into low- ($<10$~TeV) and high-energy ($>10$~TeV) events.
We chose 10~TeV as the point to split the data sets because this is the upper bin in the previous analysis~\cite{IceCube:2020phf}.
The samples have 361,633 and 6,438 events, respectively.
%The goodness-of-fit p-value for the low-energy sample is 3\% and 86\% for the high-energy sample.
%None of the nuisance parameters show differences larger than 2$\sigma$ between the best-fit points of the two samples.
Although neither result shows a closed contour, both results have best-fit points within the final sample allowed region and contours that behave as expected.
Hence, we conclude that there is no evidence of bias in the analysis due to the description of the fluxes or the energy reconstruction.

Figure~\ref{fig:threesplits} (third panel) splits the sample in the azimuthal direction, which is the angle in a horizontal plane perpendicular to the IceCube strings.
This test examines bias in the reconstruction arising from ice tilt, which is particularly interesting given recent measurements indicating not only tilt along the direction orthogonal to the ice flow but also new components not accounted for in the simulation used in this analysis~\cite{IceCube:2023qua}.
Six angles are chosen to reflect three known tilts, where three are parallel to the ice sheet and three are perpendicular.
This substantially reduces the statistics, with approximately 60,000 events in each sample.
%The goodness-of-fit p-values range from 10\% to 96\%, and the nuisance parameters agree between the different samples except for some amplitude modes of the bulk ice.
The behavior of the contours and the best-fit points is as expected.
The best-fit points scatter more than for the other cases due to the lower statistics.
This test is quite important for an analysis that searches for energy dependence associated with an oscillation and a resonance, but there is no evidence of disagreement between the split data sets and the final result.

We gave special attention to the sample split according to the zenith angle for several reasons.
First, recall that $\cos(\theta_z^{\mathrm{reco}})$ is a proxy for $L$, which rapidly changes for angles near the horizon.
This leads to a pronounced effect in signal region, as seen in the top panels of Fig.~\ref{fig:cricles}, so the description of the smearing of $\cos(\theta_z^{\nu})$ becomes very important.  
Also, above the horizon, there is a large background from downward-going muons, and while all of our studies show that these are well-removed in this analysis, a cross-check of the horizontal bins is prudent.
Lastly, as seen in Fig.~\ref{fig:cricles}, bottom right, the largest pull differences in the through-going sample, which are $\pm0.6\sigma$, are clustered in the $\cos(\theta_z^{\mathrm{reco}})>-0.05$ bin.
There are several reasons to attribute this to random effects.
First, although the pulls are among the largest, they are of both signs.
Second, the pulls are found to be randomly distributed using NIST tests~\cite{NIST}.
Third, there is no similar effect in the starting events.
Nevertheless, careful study of the signal in the horizontal versus vertical regions is warranted.  

For this cross-check, we split the data into 
different zenith directions.
The results are shown in the fourth panel of Fig.~\ref{fig:threesplits}. 
%The goodness-of-fit p-values for these samples range from 16\% (for the most horizontal region) to 79\% (for the most vertically up-going region).
%Regarding the best-fit value of the nuisance parameters, the largest discrepancy appears for the bulk ice systematics in the most horizontal sample.
We observe that none of the zenith samples show a closed contour at 95\% C.L. 
However, the best-fit points remain in the same region as those drawn from the full-zenith fit.

\begin{figure*}[t!]
    \centering
    \includegraphics[width=1.0\textwidth]{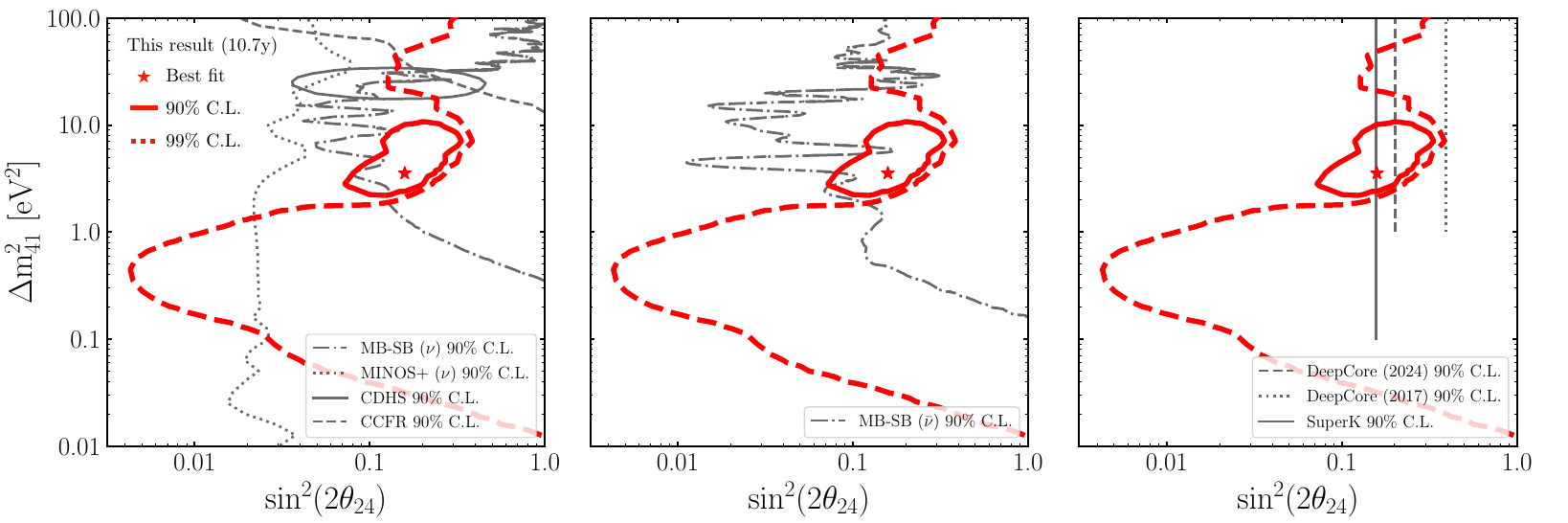}
    \caption{\textbf{Comparison with other experiments}. The 90\% and 99\% C.L. contours (red lines) compared to world data at 90\% CL. \textit{Left panel}: long-baseline experiments using $\nu_\mu$-mode beam (MiniBooNE-SciBooNE~\cite{SciBooNE:2011qyf}, MINOS~\cite{MINOS:2017cae}, CDHS and CCFR~\cite{Diaz:2019fwt}). \textit{Center panel}: long-baseline experiments using $\bar{\nu}_\mu$-mode beam (MiniBooNE-SciBooNE~\cite{MiniBooNE:2012meu}). \textit{Right panel}: atmospheric neutrino experiments (SuperK~\cite{Super-Kamiokande:2014ndf} and DeepCore~\cite{IceCube:2017ivd,Abbasi:2024ktc}).}
    \label{fig:experiments}
\end{figure*}

We also examined the influence of the prior width on the normalization, a factor that emerges as the primary systematic at higher values of $\Delta m_{41}^{2}$, as illustrated in Fig.~\ref{fig:n1_test}.
To investigate this, we reanalyzed the data with a more stringent constraint of 5\%, in contrast to the 20\% applied in the blind analysis.
The new fit prefers the same sterile neutrino hypothesis with a reduction in the null rejection by 0.3$\sigma$.
Therefore, while the uncertainty in the overall normalization contributes, the predominant driver of the null rejection is a shape-related influence.

Finally, Fig.~\ref{fig:experiments} presents this result in comparison with previous measurements of $\nu_\mu$ disappearance from other experiments. The 90\% C.L. allowed region from this result indicates an increased tension with the constraints from long-baseline experiments, namely MINOS and MiniBooNE-SciBooNE. The tension is smaller with experiments using atmospheric neutrinos.

\section{Conclusion}

This paper has presented detailed information related to the first search for a $\nu_\mu$ disappearance signature due to a 3+1 model in IceCube that separately fits 10.7 years of data divided into starting and through-going samples.
This analysis makes use of neutrino events with reconstructed energy in the 0.5 to 100~TeV region and upward going zenith angle ($\cos(\theta_z^{\mathrm{reco}})<0$).
The result is reported in Ref.~\cite{PRL}.

We have described significant improvements to the analysis since our previous search for the same signature published in 2020~\cite{IceCube:2020phf,IceCube:2020tka}.
Along with increased statistics, we have expanded the upper search range of the energy from 10~TeV to 100~TeV.
We have presented a substantial improvement to energy reconstruction, a CNN-based method of separating starting and through-going events, and an improved description of the atmospheric model.   

This paper provides a detailed list of systematic effects in the analysis.
Nuisance parameters are introduced as pull terms.
We showed that the pull term results are stable when fitting to the null (i.e., three-neutrino) and 3+1 models.

We presented an extensive discussion of the expected signature, focusing on the bin-by-bin predictions from the best fit, which was at the parameters $\sin^2(2\theta_{24}) = 0.16$ and $\Delta m^{2}_{41} = 3.5$~eV$^2$. The p-value representing the goodness of fit, found by trials, is 12\%.
Consistency with the null hypothesis of no sterile neutrino oscillations has a probability of 3.1\%. 
In this article, we have also presented a Bayesian analysis.
While the Bayesian model comparison has a different meaning than the frequentist result, the overlap of the best-fit parameter space indicates the consistency of the result even under various approaches.

After unblinding, we performed an extended set of studies where the data was split into separately fitted sub-samples.
This paper has shown that the results are stable, including for the energy and zenith angle split cases.
The reconstruction and stability test techniques reported here will also be applicable in future IceCube analyses.
These techniques are also relevant to similar studies at other neutrino telescopes.

\begin{acknowledgements}
The IceCube collaboration acknowledges the significant contributions to this manuscript from the Harvard University, Massachusetts Institute of Technology, and University of Texas at Arlington groups.
We acknowledge the support from the following agencies: USA {\textendash} U.S. National Science Foundation-Office of Polar Programs,
U.S. National Science Foundation-Physics Division,
U.S. National Science Foundation-EPSCoR,
U.S. National Science Foundation-Office of Advanced Cyberinfrastructure,
Wisconsin Alumni Research Foundation,
Center for High Throughput Computing (CHTC) at the University of Wisconsin{\textendash}Madison,
Open Science Grid (OSG),
Partnership to Advance Throughput Computing (PATh),
Advanced Cyberinfrastructure Coordination Ecosystem: Services {\&} Support (ACCESS),
Frontera computing project at the Texas Advanced Computing Center,
U.S. Department of Energy-National Energy Research Scientific Computing Center,
Particle astrophysics research computing center at the University of Maryland,
Institute for Cyber-Enabled Research at Michigan State University,
Astroparticle physics computational facility at Marquette University,
NVIDIA Corporation,
and Google Cloud Platform;
Belgium {\textendash} Funds for Scientific Research (FRS-FNRS and FWO),
FWO Odysseus and Big Science programmes,
and Belgian Federal Science Policy Office (Belspo);
Germany {\textendash} Bundesministerium f{\"u}r Bildung und Forschung (BMBF),
Deutsche Forschungsgemeinschaft (DFG),
Helmholtz Alliance for Astroparticle Physics (HAP),
Initiative and Networking Fund of the Helmholtz Association,
Deutsches Elektronen Synchrotron (DESY),
and High Performance Computing cluster of the RWTH Aachen;
Sweden {\textendash} Swedish Research Council,
Swedish Polar Research Secretariat,
Swedish National Infrastructure for Computing (SNIC),
and Knut and Alice Wallenberg Foundation;
European Union {\textendash} EGI Advanced Computing for research, and Horizon 2020 Marie Sk\l{}odowska-Curie Actions;
Australia {\textendash} Australian Research Council;
Canada {\textendash} Natural Sciences and Engineering Research Council of Canada,
Calcul Qu{\'e}bec, Compute Ontario, Canada Foundation for Innovation, WestGrid, and Digital Research Alliance of Canada;
Denmark {\textendash} Villum Fonden, Carlsberg Foundation, and European Commission;
New Zealand {\textendash} Marsden Fund;
Japan {\textendash} Japan Society for Promotion of Science (JSPS)
and Institute for Global Prominent Research (IGPR) of Chiba University;
Korea {\textendash} National Research Foundation of Korea (NRF);
Switzerland {\textendash} Swiss National Science Foundation (SNSF).
\end{acknowledgements}

\bibliographystyle{apsrev4-1}
\bibliography{bibliography}

\begin{thebibliography}{60}%
\makeatletter
\providecommand \@ifxundefined [1]{%
 \@ifx{#1\undefined}
}%
\providecommand \@ifnum [1]{%
 \ifnum #1\expandafter \@firstoftwo
 \else \expandafter \@secondoftwo
 \fi
}%
\providecommand \@ifx [1]{%
 \ifx #1\expandafter \@firstoftwo
 \else \expandafter \@secondoftwo
 \fi
}%
\providecommand \natexlab [1]{#1}%
\providecommand \enquote  [1]{``#1''}%
\providecommand \bibnamefont  [1]{#1}%
\providecommand \bibfnamefont [1]{#1}%
\providecommand \citenamefont [1]{#1}%
\providecommand \href@noop [0]{\@secondoftwo}%
\providecommand \href [0]{\begingroup \@sanitize@url \@href}%
\providecommand \@href[1]{\@@startlink{#1}\@@href}%
\providecommand \@@href[1]{\endgroup#1\@@endlink}%
\providecommand \@sanitize@url [0]{\catcode `\\12\catcode `\$12\catcode
  `\&12\catcode `\#12\catcode `\^12\catcode `\_12\catcode `\%12\relax}%
\providecommand \@@startlink[1]{}%
\providecommand \@@endlink[0]{}%
\providecommand \url  [0]{\begingroup\@sanitize@url \@url }%
\providecommand \@url [1]{\endgroup\@href {#1}{\urlprefix }}%
\providecommand \urlprefix  [0]{URL }%
\providecommand \Eprint [0]{\href }%
\providecommand \doibase [0]{http://dx.doi.org/}%
\providecommand \selectlanguage [0]{\@gobble}%
\providecommand \bibinfo  [0]{\@secondoftwo}%
\providecommand \bibfield  [0]{\@secondoftwo}%
\providecommand \translation [1]{[#1]}%
\providecommand \BibitemOpen [0]{}%
\providecommand \bibitemStop [0]{}%
\providecommand \bibitemNoStop [0]{.\EOS\space}%
\providecommand \EOS [0]{\spacefactor3000\relax}%
\providecommand \BibitemShut  [1]{\csname bibitem#1\endcsname}%
\let\auto@bib@innerbib\@empty
%</preamble>
\bibitem [{\citenamefont {Diaz}\ \emph {et~al.}(2020)\citenamefont {Diaz},
  \citenamefont {Arg\"uelles}, \citenamefont {Collin}, \citenamefont {Conrad},\
  and\ \citenamefont {Shaevitz}}]{Diaz:2019fwt}%
  \BibitemOpen
  \bibfield  {author} {\bibinfo {author} {\bibfnamefont {A.}~\bibnamefont
  {Diaz}}, \bibinfo {author} {\bibfnamefont {C.~A.}\ \bibnamefont
  {Arg\"uelles}}, \bibinfo {author} {\bibfnamefont {G.~H.}\ \bibnamefont
  {Collin}}, \bibinfo {author} {\bibfnamefont {J.~M.}\ \bibnamefont {Conrad}},
  \ and\ \bibinfo {author} {\bibfnamefont {M.~H.}\ \bibnamefont {Shaevitz}},\
  }\href {\doibase 10.1016/j.physrep.2020.08.005} {\bibfield  {journal}
  {\bibinfo  {journal} {Phys. Rept.}\ }\textbf {\bibinfo {volume} {884}},\
  \bibinfo {pages} {1} (\bibinfo {year} {2020})},\ \Eprint
  {http://arxiv.org/abs/1906.00045} {arXiv:1906.00045 [hep-ex]} \BibitemShut
  {NoStop}%
\bibitem [{\citenamefont {Dentler}\ \emph {et~al.}(2018)\citenamefont
  {Dentler}, \citenamefont {Hern\'andez-Cabezudo}, \citenamefont {Kopp},
  \citenamefont {Machado}, \citenamefont {Maltoni}, \citenamefont
  {Martinez-Soler},\ and\ \citenamefont {Schwetz}}]{Dentler:2018sju}%
  \BibitemOpen
  \bibfield  {author} {\bibinfo {author} {\bibfnamefont {M.}~\bibnamefont
  {Dentler}}, \bibinfo {author} {\bibfnamefont {A.}~\bibnamefont
  {Hern\'andez-Cabezudo}}, \bibinfo {author} {\bibfnamefont {J.}~\bibnamefont
  {Kopp}}, \bibinfo {author} {\bibfnamefont {P.~A.~N.}\ \bibnamefont
  {Machado}}, \bibinfo {author} {\bibfnamefont {M.}~\bibnamefont {Maltoni}},
  \bibinfo {author} {\bibfnamefont {I.}~\bibnamefont {Martinez-Soler}}, \ and\
  \bibinfo {author} {\bibfnamefont {T.}~\bibnamefont {Schwetz}},\ }\href
  {\doibase 10.1007/JHEP08(2018)010} {\bibfield  {journal} {\bibinfo  {journal}
  {JHEP}\ }\textbf {\bibinfo {volume} {08}},\ \bibinfo {pages} {010} (\bibinfo
  {year} {2018})},\ \Eprint {http://arxiv.org/abs/1803.10661} {arXiv:1803.10661
  [hep-ph]} \BibitemShut {NoStop}%
\bibitem [{\citenamefont {Gariazzo}\ \emph {et~al.}(2017)\citenamefont
  {Gariazzo}, \citenamefont {Giunti}, \citenamefont {Laveder},\ and\
  \citenamefont {Li}}]{Gariazzo:2017fdh}%
  \BibitemOpen
  \bibfield  {author} {\bibinfo {author} {\bibfnamefont {S.}~\bibnamefont
  {Gariazzo}}, \bibinfo {author} {\bibfnamefont {C.}~\bibnamefont {Giunti}},
  \bibinfo {author} {\bibfnamefont {M.}~\bibnamefont {Laveder}}, \ and\
  \bibinfo {author} {\bibfnamefont {Y.~F.}\ \bibnamefont {Li}},\ }\href
  {\doibase 10.1007/JHEP06(2017)135} {\bibfield  {journal} {\bibinfo  {journal}
  {JHEP}\ }\textbf {\bibinfo {volume} {06}},\ \bibinfo {pages} {135} (\bibinfo
  {year} {2017})},\ \Eprint {http://arxiv.org/abs/1703.00860} {arXiv:1703.00860
  [hep-ph]} \BibitemShut {NoStop}%
%%CITATION = ARXIV:1703.00860;%%
\bibitem [{\citenamefont {Athanassopoulos}\ \emph {et~al.}(1996)\citenamefont
  {Athanassopoulos} \emph {et~al.}}]{PhysRevLett.77.3082}%
  \BibitemOpen
  \bibfield  {author} {\bibinfo {author} {\bibfnamefont {C.}~\bibnamefont
  {Athanassopoulos}} \emph {et~al.} (\bibinfo {collaboration} {LSND
  Collaboration}),\ }\href {\doibase 10.1103/PhysRevLett.77.3082} {\bibfield
  {journal} {\bibinfo  {journal} {Phys. Rev. Lett.}\ }\textbf {\bibinfo
  {volume} {77}},\ \bibinfo {pages} {3082} (\bibinfo {year}
  {1996})}\BibitemShut {NoStop}%
\bibitem [{\citenamefont {Aguilar-Arevalo}\ \emph {et~al.}(2013)\citenamefont
  {Aguilar-Arevalo} \emph {et~al.}}]{PhysRevLett.110.161801}%
  \BibitemOpen
  \bibfield  {author} {\bibinfo {author} {\bibfnamefont {A.~A.}\ \bibnamefont
  {Aguilar-Arevalo}} \emph {et~al.} (\bibinfo {collaboration} {MiniBooNE
  Collaboration}),\ }\href {\doibase 10.1103/PhysRevLett.110.161801} {\bibfield
   {journal} {\bibinfo  {journal} {Phys. Rev. Lett.}\ }\textbf {\bibinfo
  {volume} {110}},\ \bibinfo {pages} {161801} (\bibinfo {year}
  {2013})}\BibitemShut {NoStop}%
\bibitem [{\citenamefont {Barinov}\ \emph {et~al.}(2022)\citenamefont {Barinov}
  \emph {et~al.}}]{PhysRevLett.128.232501}%
  \BibitemOpen
  \bibfield  {author} {\bibinfo {author} {\bibfnamefont {V.~V.}\ \bibnamefont
  {Barinov}} \emph {et~al.},\ }\href {\doibase 10.1103/PhysRevLett.128.232501}
  {\bibfield  {journal} {\bibinfo  {journal} {Phys. Rev. Lett.}\ }\textbf
  {\bibinfo {volume} {128}},\ \bibinfo {pages} {232501} (\bibinfo {year}
  {2022})}\BibitemShut {NoStop}%
\bibitem [{\citenamefont {for PRL~reference}()}]{PRL}%
  \BibitemOpen
  \bibfield  {author} {\bibinfo {author} {\bibfnamefont {P.}~\bibnamefont {for
  PRL~reference}},\ }\href@noop {} {\ }\BibitemShut {NoStop}%
\bibitem [{\citenamefont {Aartsen}\ \emph
  {et~al.}(2017{\natexlab{a}})\citenamefont {Aartsen} \emph
  {et~al.}}]{Aartsen:2016nxy}%
  \BibitemOpen
  \bibfield  {author} {\bibinfo {author} {\bibfnamefont {M.~G.}\ \bibnamefont
  {Aartsen}} \emph {et~al.} (\bibinfo {collaboration} {IceCube}),\ }\href
  {\doibase 10.1088/1748-0221/12/03/P03012} {\bibfield  {journal} {\bibinfo
  {journal} {JINST}\ }\textbf {\bibinfo {volume} {12}},\ \bibinfo {pages}
  {P03012} (\bibinfo {year} {2017}{\natexlab{a}})},\ \Eprint
  {http://arxiv.org/abs/1612.05093} {arXiv:1612.05093 [astro-ph.IM]}
  \BibitemShut {NoStop}%
%%CITATION = ARXIV:1612.05093;%%
\bibitem [{\citenamefont {Abbasi}\ \emph
  {et~al.}(2009{\natexlab{a}})\citenamefont {Abbasi} \emph
  {et~al.}}]{IceCube:2008qbc}%
  \BibitemOpen
  \bibfield  {author} {\bibinfo {author} {\bibfnamefont {R.}~\bibnamefont
  {Abbasi}} \emph {et~al.} (\bibinfo {collaboration} {IceCube}),\ }\href
  {\doibase 10.1016/j.nima.2009.01.001} {\bibfield  {journal} {\bibinfo
  {journal} {Nucl. Instrum. Meth. A}\ }\textbf {\bibinfo {volume} {601}},\
  \bibinfo {pages} {294} (\bibinfo {year} {2009}{\natexlab{a}})},\ \Eprint
  {http://arxiv.org/abs/0810.4930} {arXiv:0810.4930 [physics.ins-det]}
  \BibitemShut {NoStop}%
\bibitem [{\citenamefont {Abbasi}\ \emph {et~al.}(2012)\citenamefont {Abbasi}
  \emph {et~al.}}]{IceCube:2011ucd}%
  \BibitemOpen
  \bibfield  {author} {\bibinfo {author} {\bibfnamefont {R.}~\bibnamefont
  {Abbasi}} \emph {et~al.} (\bibinfo {collaboration} {IceCube}),\ }\href
  {\doibase 10.1016/j.astropartphys.2012.01.004} {\bibfield  {journal}
  {\bibinfo  {journal} {Astropart. Phys.}\ }\textbf {\bibinfo {volume} {35}},\
  \bibinfo {pages} {615} (\bibinfo {year} {2012})},\ \Eprint
  {http://arxiv.org/abs/1109.6096} {arXiv:1109.6096 [astro-ph.IM]} \BibitemShut
  {NoStop}%
\bibitem [{\citenamefont {Barr}\ \emph {et~al.}(2006)\citenamefont {Barr},
  \citenamefont {Gaisser}, \citenamefont {Robbins},\ and\ \citenamefont
  {Stanev}}]{Barr:2006it}%
  \BibitemOpen
  \bibfield  {author} {\bibinfo {author} {\bibfnamefont {G.~D.}\ \bibnamefont
  {Barr}}, \bibinfo {author} {\bibfnamefont {T.~K.}\ \bibnamefont {Gaisser}},
  \bibinfo {author} {\bibfnamefont {S.}~\bibnamefont {Robbins}}, \ and\
  \bibinfo {author} {\bibfnamefont {T.}~\bibnamefont {Stanev}},\ }\href
  {\doibase 10.1103/PhysRevD.74.094009} {\bibfield  {journal} {\bibinfo
  {journal} {Phys. Rev. D}\ }\textbf {\bibinfo {volume} {74}},\ \bibinfo
  {pages} {094009} (\bibinfo {year} {2006})},\ \Eprint
  {http://arxiv.org/abs/astro-ph/0611266} {arXiv:astro-ph/0611266} \BibitemShut
  {NoStop}%
\bibitem [{\citenamefont {Fedynitch}\ and\ \citenamefont
  {Huber}(2022)}]{Fedynitch:2022vty}%
  \BibitemOpen
  \bibfield  {author} {\bibinfo {author} {\bibfnamefont {A.}~\bibnamefont
  {Fedynitch}}\ and\ \bibinfo {author} {\bibfnamefont {M.}~\bibnamefont
  {Huber}},\ }\href {\doibase 10.1103/PhysRevD.106.083018} {\bibfield
  {journal} {\bibinfo  {journal} {Phys. Rev. D}\ }\textbf {\bibinfo {volume}
  {106}},\ \bibinfo {pages} {083018} (\bibinfo {year} {2022})},\ \Eprint
  {http://arxiv.org/abs/2205.14766} {arXiv:2205.14766 [astro-ph.HE]}
  \BibitemShut {NoStop}%
\bibitem [{\citenamefont {Ya\~nez}\ and\ \citenamefont
  {Fedynitch}(2023)}]{Yanez:2023lsy}%
  \BibitemOpen
  \bibfield  {author} {\bibinfo {author} {\bibfnamefont {J.~P.}\ \bibnamefont
  {Ya\~nez}}\ and\ \bibinfo {author} {\bibfnamefont {A.}~\bibnamefont
  {Fedynitch}},\ }\href {\doibase 10.1103/PhysRevD.107.123037} {\bibfield
  {journal} {\bibinfo  {journal} {Phys. Rev. D}\ }\textbf {\bibinfo {volume}
  {107}},\ \bibinfo {pages} {123037} (\bibinfo {year} {2023})},\ \Eprint
  {http://arxiv.org/abs/2303.00022} {arXiv:2303.00022 [hep-ph]} \BibitemShut
  {NoStop}%
\bibitem [{\citenamefont {Abbasi}\ \emph
  {et~al.}(2021{\natexlab{a}})\citenamefont {Abbasi} \emph
  {et~al.}}]{IceCube:2020tcq}%
  \BibitemOpen
  \bibfield  {author} {\bibinfo {author} {\bibfnamefont {R.}~\bibnamefont
  {Abbasi}} \emph {et~al.} (\bibinfo {collaboration} {IceCube}),\ }\href
  {\doibase 10.1016/j.cpc.2021.108018} {\bibfield  {journal} {\bibinfo
  {journal} {Comput. Phys. Commun.}\ }\textbf {\bibinfo {volume} {266}},\
  \bibinfo {pages} {108018} (\bibinfo {year} {2021}{\natexlab{a}})},\ \Eprint
  {http://arxiv.org/abs/2012.10449} {arXiv:2012.10449 [physics.comp-ph]}
  \BibitemShut {NoStop}%
\bibitem [{\citenamefont {Abbasi}\ \emph
  {et~al.}(2009{\natexlab{b}})\citenamefont {Abbasi} \emph
  {et~al.}}]{Abbasi:2008aa}%
  \BibitemOpen
  \bibfield  {author} {\bibinfo {author} {\bibfnamefont {R.}~\bibnamefont
  {Abbasi}} \emph {et~al.} (\bibinfo {collaboration} {IceCube}),\ }\href
  {\doibase 10.1016/j.nima.2009.01.001} {\bibfield  {journal} {\bibinfo
  {journal} {Nucl. Instrum. Meth.}\ }\textbf {\bibinfo {volume} {A601}},\
  \bibinfo {pages} {294} (\bibinfo {year} {2009}{\natexlab{b}})},\ \Eprint
  {http://arxiv.org/abs/0810.4930} {arXiv:0810.4930 [physics.ins-det]}
  \BibitemShut {NoStop}%
%%CITATION = ARXIV:0810.4930;%%
\bibitem [{\citenamefont {Rongen}(2019)}]{Rongen:2019wsh}%
  \BibitemOpen
  \bibfield  {author} {\bibinfo {author} {\bibfnamefont {M.}~\bibnamefont
  {Rongen}},\ }\emph {\bibinfo {title} {{Calibration of the IceCube Neutrino
  Observatory}}},\ \href {\doibase 10.18154/RWTH-2019-09941} {Ph.D. thesis},\
  \bibinfo  {school} {RWTH Aachen U.} (\bibinfo {year} {2019}),\ \Eprint
  {http://arxiv.org/abs/1911.02016} {arXiv:1911.02016 [astro-ph.IM]}
  \BibitemShut {NoStop}%
\bibitem [{\citenamefont {Aartsen}\ \emph
  {et~al.}(2020{\natexlab{a}})\citenamefont {Aartsen} \emph
  {et~al.}}]{IceCube:2020phf}%
  \BibitemOpen
  \bibfield  {author} {\bibinfo {author} {\bibfnamefont {M.~G.}\ \bibnamefont
  {Aartsen}} \emph {et~al.} (\bibinfo {collaboration} {IceCube}),\ }\href
  {\doibase 10.1103/PhysRevLett.125.141801} {\bibfield  {journal} {\bibinfo
  {journal} {Phys. Rev. Lett.}\ }\textbf {\bibinfo {volume} {125}},\ \bibinfo
  {pages} {141801} (\bibinfo {year} {2020}{\natexlab{a}})},\ \Eprint
  {http://arxiv.org/abs/2005.12942} {arXiv:2005.12942 [hep-ex]} \BibitemShut
  {NoStop}%
\bibitem [{\citenamefont {Ahrens}\ \emph {et~al.}(2004)\citenamefont {Ahrens}
  \emph {et~al.}}]{AMANDA:2003vtt}%
  \BibitemOpen
  \bibfield  {author} {\bibinfo {author} {\bibfnamefont {J.}~\bibnamefont
  {Ahrens}} \emph {et~al.} (\bibinfo {collaboration} {AMANDA}),\ }\href
  {\doibase 10.1016/j.nima.2004.01.065} {\bibfield  {journal} {\bibinfo
  {journal} {Nucl. Instrum. Meth. A}\ }\textbf {\bibinfo {volume} {524}},\
  \bibinfo {pages} {169} (\bibinfo {year} {2004})},\ \Eprint
  {http://arxiv.org/abs/astro-ph/0407044} {arXiv:astro-ph/0407044} \BibitemShut
  {NoStop}%
\bibitem [{\citenamefont {Aartsen}\ \emph
  {et~al.}(2014{\natexlab{a}})\citenamefont {Aartsen} \emph
  {et~al.}}]{Aartsen:2013bfa}%
  \BibitemOpen
  \bibfield  {author} {\bibinfo {author} {\bibfnamefont {M.~G.}\ \bibnamefont
  {Aartsen}} \emph {et~al.},\ }\href {\doibase 10.1016/j.nima.2013.10.074}
  {\bibfield  {journal} {\bibinfo  {journal} {Nucl. Instrum. Meth. A}\ }\textbf
  {\bibinfo {volume} {736}},\ \bibinfo {pages} {143} (\bibinfo {year}
  {2014}{\natexlab{a}})},\ \Eprint {http://arxiv.org/abs/1308.5501}
  {arXiv:1308.5501 [astro-ph.IM]} \BibitemShut {NoStop}%
\bibitem [{\citenamefont {Aartsen}\ \emph
  {et~al.}(2020{\natexlab{b}})\citenamefont {Aartsen} \emph
  {et~al.}}]{IceCube:2020tka}%
  \BibitemOpen
  \bibfield  {author} {\bibinfo {author} {\bibfnamefont {M.~G.}\ \bibnamefont
  {Aartsen}} \emph {et~al.} (\bibinfo {collaboration} {IceCube}),\ }\href
  {\doibase 10.1103/PhysRevD.102.052009} {\bibfield  {journal} {\bibinfo
  {journal} {Phys. Rev. D}\ }\textbf {\bibinfo {volume} {102}},\ \bibinfo
  {pages} {052009} (\bibinfo {year} {2020}{\natexlab{b}})},\ \Eprint
  {http://arxiv.org/abs/2005.12943} {arXiv:2005.12943 [hep-ex]} \BibitemShut
  {NoStop}%
\bibitem [{\citenamefont {Hastie}\ \emph {et~al.}(2009)\citenamefont {Hastie},
  \citenamefont {Rosset}, \citenamefont {Zhu},\ and\ \citenamefont
  {Zou}}]{Hastie2009MulticlassA}%
  \BibitemOpen
  \bibfield  {author} {\bibinfo {author} {\bibfnamefont {T.~J.}\ \bibnamefont
  {Hastie}}, \bibinfo {author} {\bibfnamefont {S.}~\bibnamefont {Rosset}},
  \bibinfo {author} {\bibfnamefont {J.}~\bibnamefont {Zhu}}, \ and\ \bibinfo
  {author} {\bibfnamefont {H.}~\bibnamefont {Zou}},\ }\href
  {https://api.semanticscholar.org/CorpusID:11803458} {\bibfield  {journal}
  {\bibinfo  {journal} {Statistics and Its Interface}\ }\textbf {\bibinfo
  {volume} {2}},\ \bibinfo {pages} {349} (\bibinfo {year} {2009})}\BibitemShut
  {NoStop}%
\bibitem [{\citenamefont {Aartsen}\ \emph
  {et~al.}(2014{\natexlab{b}})\citenamefont {Aartsen} \emph
  {et~al.}}]{Aartsen:2013vja}%
  \BibitemOpen
  \bibfield  {author} {\bibinfo {author} {\bibfnamefont {M.~G.}\ \bibnamefont
  {Aartsen}} \emph {et~al.} (\bibinfo {collaboration} {IceCube}),\ }\href
  {\doibase 10.1088/1748-0221/9/03/P03009} {\bibfield  {journal} {\bibinfo
  {journal} {JINST}\ }\textbf {\bibinfo {volume} {9}},\ \bibinfo {pages}
  {P03009} (\bibinfo {year} {2014}{\natexlab{b}})},\ \Eprint
  {http://arxiv.org/abs/1311.4767} {arXiv:1311.4767 [physics.ins-det]}
  \BibitemShut {NoStop}%
%%CITATION = ARXIV:1311.4767;%%
\bibitem [{\citenamefont {Abbasi}\ \emph
  {et~al.}(2021{\natexlab{b}})\citenamefont {Abbasi} \emph
  {et~al.}}]{Abbasi:2021ryj}%
  \BibitemOpen
  \bibfield  {author} {\bibinfo {author} {\bibfnamefont {R.}~\bibnamefont
  {Abbasi}} \emph {et~al.},\ }\href {\doibase 10.1088/1748-0221/16/07/P07041}
  {\bibfield  {journal} {\bibinfo  {journal} {JINST}\ }\textbf {\bibinfo
  {volume} {16}},\ \bibinfo {pages} {P07041} (\bibinfo {year}
  {2021}{\natexlab{b}})},\ \Eprint {http://arxiv.org/abs/2101.11589}
  {arXiv:2101.11589 [hep-ex]} \BibitemShut {NoStop}%
\bibitem [{\citenamefont {Abbasi}\ \emph
  {et~al.}(2023{\natexlab{a}})\citenamefont {Abbasi} \emph
  {et~al.}}]{IceCube:2023ame}%
  \BibitemOpen
  \bibfield  {author} {\bibinfo {author} {\bibfnamefont {R.}~\bibnamefont
  {Abbasi}} \emph {et~al.} (\bibinfo {collaboration} {IceCube}),\ }\href
  {\doibase 10.1126/science.adc9818} {\bibfield  {journal} {\bibinfo  {journal}
  {Science}\ }\textbf {\bibinfo {volume} {380}},\ \bibinfo {pages} {adc9818}
  (\bibinfo {year} {2023}{\natexlab{a}})},\ \Eprint
  {http://arxiv.org/abs/2307.04427} {arXiv:2307.04427 [astro-ph.HE]}
  \BibitemShut {NoStop}%
\bibitem [{\citenamefont {Radel}\ and\ \citenamefont
  {Wiebusch}(2012)}]{Radel:2012kw}%
  \BibitemOpen
  \bibfield  {author} {\bibinfo {author} {\bibfnamefont {L.}~\bibnamefont
  {Radel}}\ and\ \bibinfo {author} {\bibfnamefont {C.}~\bibnamefont
  {Wiebusch}},\ }\href {\doibase 10.1016/j.astropartphys.2012.09.008}
  {\bibfield  {journal} {\bibinfo  {journal} {Astropart. Phys.}\ }\textbf
  {\bibinfo {volume} {38}},\ \bibinfo {pages} {53} (\bibinfo {year} {2012})},\
  \Eprint {http://arxiv.org/abs/1206.5530} {arXiv:1206.5530 [astro-ph.IM]}
  \BibitemShut {NoStop}%
\bibitem [{\citenamefont {Kronmueller}\ and\ \citenamefont
  {Glauch}(2020)}]{Kronmueller:2019jzh}%
  \BibitemOpen
  \bibfield  {author} {\bibinfo {author} {\bibfnamefont {M.}~\bibnamefont
  {Kronmueller}}\ and\ \bibinfo {author} {\bibfnamefont {T.}~\bibnamefont
  {Glauch}} (\bibinfo {collaboration} {IceCube}),\ }\href {\doibase
  10.22323/1.358.0937} {\bibfield  {journal} {\bibinfo  {journal} {PoS}\
  }\textbf {\bibinfo {volume} {ICRC2019}},\ \bibinfo {pages} {937} (\bibinfo
  {year} {2020})},\ \Eprint {http://arxiv.org/abs/1908.08763} {arXiv:1908.08763
  [astro-ph.IM]} \BibitemShut {NoStop}%
\bibitem [{\citenamefont {Riehn}\ \emph {et~al.}(2018)\citenamefont {Riehn},
  \citenamefont {Dembinski}, \citenamefont {Engel}, \citenamefont {Fedynitch},
  \citenamefont {Gaisser},\ and\ \citenamefont {Stanev}}]{Riehn:2017mfm}%
  \BibitemOpen
  \bibfield  {author} {\bibinfo {author} {\bibfnamefont {F.}~\bibnamefont
  {Riehn}}, \bibinfo {author} {\bibfnamefont {H.~P.}\ \bibnamefont
  {Dembinski}}, \bibinfo {author} {\bibfnamefont {R.}~\bibnamefont {Engel}},
  \bibinfo {author} {\bibfnamefont {A.}~\bibnamefont {Fedynitch}}, \bibinfo
  {author} {\bibfnamefont {T.~K.}\ \bibnamefont {Gaisser}}, \ and\ \bibinfo
  {author} {\bibfnamefont {T.}~\bibnamefont {Stanev}},\ }\href {\doibase
  10.22323/1.301.0301} {\bibfield  {journal} {\bibinfo  {journal} {PoS}\
  }\textbf {\bibinfo {volume} {ICRC2017}},\ \bibinfo {pages} {301} (\bibinfo
  {year} {2018})},\ \Eprint {http://arxiv.org/abs/1709.07227} {arXiv:1709.07227
  [hep-ph]} \BibitemShut {NoStop}%
\bibitem [{\citenamefont {Fedynitch}\ \emph {et~al.}(2019)\citenamefont
  {Fedynitch}, \citenamefont {Riehn}, \citenamefont {Engel}, \citenamefont
  {Gaisser},\ and\ \citenamefont {Stanev}}]{Fedynitch:2018cbl}%
  \BibitemOpen
  \bibfield  {author} {\bibinfo {author} {\bibfnamefont {A.}~\bibnamefont
  {Fedynitch}}, \bibinfo {author} {\bibfnamefont {F.}~\bibnamefont {Riehn}},
  \bibinfo {author} {\bibfnamefont {R.}~\bibnamefont {Engel}}, \bibinfo
  {author} {\bibfnamefont {T.~K.}\ \bibnamefont {Gaisser}}, \ and\ \bibinfo
  {author} {\bibfnamefont {T.}~\bibnamefont {Stanev}},\ }\href {\doibase
  10.1103/PhysRevD.100.103018} {\bibfield  {journal} {\bibinfo  {journal}
  {Phys. Rev. D}\ }\textbf {\bibinfo {volume} {100}},\ \bibinfo {pages}
  {103018} (\bibinfo {year} {2019})},\ \Eprint
  {http://arxiv.org/abs/1806.04140} {arXiv:1806.04140 [hep-ph]} \BibitemShut
  {NoStop}%
\bibitem [{\citenamefont {Abbasi}\ \emph
  {et~al.}(2021{\natexlab{c}})\citenamefont {Abbasi} \emph
  {et~al.}}]{IceCube:2020wum}%
  \BibitemOpen
  \bibfield  {author} {\bibinfo {author} {\bibfnamefont {R.}~\bibnamefont
  {Abbasi}} \emph {et~al.} (\bibinfo {collaboration} {IceCube}),\ }\href
  {\doibase 10.1103/PhysRevD.104.022002} {\bibfield  {journal} {\bibinfo
  {journal} {Phys. Rev. D}\ }\textbf {\bibinfo {volume} {104}},\ \bibinfo
  {pages} {022002} (\bibinfo {year} {2021}{\natexlab{c}})},\ \Eprint
  {http://arxiv.org/abs/2011.03545} {arXiv:2011.03545 [astro-ph.HE]}
  \BibitemShut {NoStop}%
\bibitem [{\citenamefont {Abbasi}\ \emph {et~al.}(2022)\citenamefont {Abbasi}
  \emph {et~al.}}]{IceCube:2021uhz}%
  \BibitemOpen
  \bibfield  {author} {\bibinfo {author} {\bibfnamefont {R.}~\bibnamefont
  {Abbasi}} \emph {et~al.} (\bibinfo {collaboration} {IceCube}),\ }\href
  {\doibase 10.3847/1538-4357/ac4d29} {\bibfield  {journal} {\bibinfo
  {journal} {Astrophys. J.}\ }\textbf {\bibinfo {volume} {928}},\ \bibinfo
  {pages} {50} (\bibinfo {year} {2022})},\ \Eprint
  {http://arxiv.org/abs/2111.10299} {arXiv:2111.10299 [astro-ph.HE]}
  \BibitemShut {NoStop}%
\bibitem [{\citenamefont {Aartsen}\ \emph
  {et~al.}(2020{\natexlab{c}})\citenamefont {Aartsen} \emph
  {et~al.}}]{IceCube:2020acn}%
  \BibitemOpen
  \bibfield  {author} {\bibinfo {author} {\bibfnamefont {M.~G.}\ \bibnamefont
  {Aartsen}} \emph {et~al.} (\bibinfo {collaboration} {IceCube}),\ }\href
  {\doibase 10.1103/PhysRevLett.125.121104} {\bibfield  {journal} {\bibinfo
  {journal} {Phys. Rev. Lett.}\ }\textbf {\bibinfo {volume} {125}},\ \bibinfo
  {pages} {121104} (\bibinfo {year} {2020}{\natexlab{c}})},\ \Eprint
  {http://arxiv.org/abs/2001.09520} {arXiv:2001.09520 [astro-ph.HE]}
  \BibitemShut {NoStop}%
\bibitem [{\citenamefont {Cooper-Sarkar}\ \emph {et~al.}(2011)\citenamefont
  {Cooper-Sarkar}, \citenamefont {Mertsch},\ and\ \citenamefont
  {Sarkar}}]{Cooper-Sarkar:2011jtt}%
  \BibitemOpen
  \bibfield  {author} {\bibinfo {author} {\bibfnamefont {A.}~\bibnamefont
  {Cooper-Sarkar}}, \bibinfo {author} {\bibfnamefont {P.}~\bibnamefont
  {Mertsch}}, \ and\ \bibinfo {author} {\bibfnamefont {S.}~\bibnamefont
  {Sarkar}},\ }\href {\doibase 10.1007/JHEP08(2011)042} {\bibfield  {journal}
  {\bibinfo  {journal} {JHEP}\ }\textbf {\bibinfo {volume} {08}},\ \bibinfo
  {pages} {042} (\bibinfo {year} {2011})},\ \Eprint
  {http://arxiv.org/abs/1106.3723} {arXiv:1106.3723 [hep-ph]} \BibitemShut
  {NoStop}%
\bibitem [{\citenamefont {Binder}(2017)}]{Binder:2017rlx}%
  \BibitemOpen
  \bibfield  {author} {\bibinfo {author} {\bibfnamefont {G.~A.}\ \bibnamefont
  {Binder}},\ }\emph {\bibinfo {title} {{Measurements of the Flavor Composition
  and Inelasticity Distribution of High-Energy Neutrino Interactions in
  IceCube}}},\ \href@noop {} {Ph.D. thesis},\ \bibinfo  {school} {UC, Berkeley}
  (\bibinfo {year} {2017})\BibitemShut {NoStop}%
\bibitem [{\citenamefont {Klein}\ \emph {et~al.}(2020)\citenamefont {Klein},
  \citenamefont {Robertson},\ and\ \citenamefont {Vogt}}]{Klein:2020nuk}%
  \BibitemOpen
  \bibfield  {author} {\bibinfo {author} {\bibfnamefont {S.~R.}\ \bibnamefont
  {Klein}}, \bibinfo {author} {\bibfnamefont {S.~A.}\ \bibnamefont
  {Robertson}}, \ and\ \bibinfo {author} {\bibfnamefont {R.}~\bibnamefont
  {Vogt}},\ }\href {\doibase 10.1103/PhysRevC.102.015808} {\bibfield  {journal}
  {\bibinfo  {journal} {Phys. Rev. C}\ }\textbf {\bibinfo {volume} {102}},\
  \bibinfo {pages} {015808} (\bibinfo {year} {2020})},\ \Eprint
  {http://arxiv.org/abs/2001.03677} {arXiv:2001.03677 [hep-ph]} \BibitemShut
  {NoStop}%
\bibitem [{\citenamefont {Garcia}\ \emph {et~al.}(2020)\citenamefont {Garcia},
  \citenamefont {Gauld}, \citenamefont {Heijboer},\ and\ \citenamefont
  {Rojo}}]{Garcia:2020jwr}%
  \BibitemOpen
  \bibfield  {author} {\bibinfo {author} {\bibfnamefont {A.}~\bibnamefont
  {Garcia}}, \bibinfo {author} {\bibfnamefont {R.}~\bibnamefont {Gauld}},
  \bibinfo {author} {\bibfnamefont {A.}~\bibnamefont {Heijboer}}, \ and\
  \bibinfo {author} {\bibfnamefont {J.}~\bibnamefont {Rojo}},\ }\href {\doibase
  10.1088/1475-7516/2020/09/025} {\bibfield  {journal} {\bibinfo  {journal}
  {JCAP}\ }\textbf {\bibinfo {volume} {09}},\ \bibinfo {pages} {025} (\bibinfo
  {year} {2020})},\ \Eprint {http://arxiv.org/abs/2004.04756} {arXiv:2004.04756
  [hep-ph]} \BibitemShut {NoStop}%
\bibitem [{\citenamefont {Xie}\ \emph {et~al.}(2023)\citenamefont {Xie},
  \citenamefont {Gao}, \citenamefont {Hobbs}, \citenamefont {Stump},\ and\
  \citenamefont {Yuan}}]{Xie:2023suk}%
  \BibitemOpen
  \bibfield  {author} {\bibinfo {author} {\bibfnamefont {K.}~\bibnamefont
  {Xie}}, \bibinfo {author} {\bibfnamefont {J.}~\bibnamefont {Gao}}, \bibinfo
  {author} {\bibfnamefont {T.~J.}\ \bibnamefont {Hobbs}}, \bibinfo {author}
  {\bibfnamefont {D.~R.}\ \bibnamefont {Stump}}, \ and\ \bibinfo {author}
  {\bibfnamefont {C.~P.}\ \bibnamefont {Yuan}},\ }\href@noop {} {\  (\bibinfo
  {year} {2023})},\ \Eprint {http://arxiv.org/abs/2303.13607} {arXiv:2303.13607
  [hep-ph]} \BibitemShut {NoStop}%
\bibitem [{\citenamefont {Jeong}\ and\ \citenamefont
  {Reno}(2023)}]{Jeong:2023hwe}%
  \BibitemOpen
  \bibfield  {author} {\bibinfo {author} {\bibfnamefont {Y.~S.}\ \bibnamefont
  {Jeong}}\ and\ \bibinfo {author} {\bibfnamefont {M.~H.}\ \bibnamefont
  {Reno}},\ }\href {\doibase 10.1103/PhysRevD.108.113010} {\bibfield  {journal}
  {\bibinfo  {journal} {Phys. Rev. D}\ }\textbf {\bibinfo {volume} {108}},\
  \bibinfo {pages} {113010} (\bibinfo {year} {2023})},\ \Eprint
  {http://arxiv.org/abs/2307.09241} {arXiv:2307.09241 [hep-ph]} \BibitemShut
  {NoStop}%
\bibitem [{\citenamefont {Candido}\ \emph {et~al.}(2023)\citenamefont
  {Candido}, \citenamefont {Garcia}, \citenamefont {Magni}, \citenamefont
  {Rabemananjara}, \citenamefont {Rojo},\ and\ \citenamefont
  {Stegeman}}]{Candido:2023utz}%
  \BibitemOpen
  \bibfield  {author} {\bibinfo {author} {\bibfnamefont {A.}~\bibnamefont
  {Candido}}, \bibinfo {author} {\bibfnamefont {A.}~\bibnamefont {Garcia}},
  \bibinfo {author} {\bibfnamefont {G.}~\bibnamefont {Magni}}, \bibinfo
  {author} {\bibfnamefont {T.}~\bibnamefont {Rabemananjara}}, \bibinfo {author}
  {\bibfnamefont {J.}~\bibnamefont {Rojo}}, \ and\ \bibinfo {author}
  {\bibfnamefont {R.}~\bibnamefont {Stegeman}},\ }\href {\doibase
  10.1007/JHEP05(2023)149} {\bibfield  {journal} {\bibinfo  {journal} {JHEP}\
  }\textbf {\bibinfo {volume} {05}},\ \bibinfo {pages} {149} (\bibinfo {year}
  {2023})},\ \Eprint {http://arxiv.org/abs/2302.08527} {arXiv:2302.08527
  [hep-ph]} \BibitemShut {NoStop}%
\bibitem [{\citenamefont {Reno}(2023)}]{Reno:2023sdm}%
  \BibitemOpen
  \bibfield  {author} {\bibinfo {author} {\bibfnamefont {M.~H.}\ \bibnamefont
  {Reno}},\ }\href {\doibase 10.1146/annurev-nucl-111422-040200} {\bibfield
  {journal} {\bibinfo  {journal} {Ann. Rev. Nucl. Part. Sci.}\ }\textbf
  {\bibinfo {volume} {73}},\ \bibinfo {pages} {181} (\bibinfo {year}
  {2023})}\BibitemShut {NoStop}%
\bibitem [{\citenamefont {Sandrock}\ \emph {et~al.}(2020)\citenamefont
  {Sandrock}, \citenamefont {Kokoulin},\ and\ \citenamefont
  {Petrukhin}}]{Sandrock_2020}%
  \BibitemOpen
  \bibfield  {author} {\bibinfo {author} {\bibfnamefont {A.}~\bibnamefont
  {Sandrock}}, \bibinfo {author} {\bibfnamefont {R.~P.}\ \bibnamefont
  {Kokoulin}}, \ and\ \bibinfo {author} {\bibfnamefont {A.~A.}\ \bibnamefont
  {Petrukhin}},\ }\href {\doibase 10.1088/1742-6596/1690/1/012005} {\bibfield
  {journal} {\bibinfo  {journal} {Journal of Physics: Conference Series}\
  }\textbf {\bibinfo {volume} {1690}},\ \bibinfo {pages} {012005} (\bibinfo
  {year} {2020})}\BibitemShut {NoStop}%
\bibitem [{\citenamefont {Plestid}\ and\ \citenamefont
  {Zhou}(2024)}]{Plestid:2024bva}%
  \BibitemOpen
  \bibfield  {author} {\bibinfo {author} {\bibfnamefont {R.}~\bibnamefont
  {Plestid}}\ and\ \bibinfo {author} {\bibfnamefont {B.}~\bibnamefont {Zhou}},\
  }\href@noop {} {\  (\bibinfo {year} {2024})},\ \Eprint
  {http://arxiv.org/abs/2403.07984} {arXiv:2403.07984 [hep-ph]} \BibitemShut
  {NoStop}%
\bibitem [{\citenamefont {Kennett}\ \emph {et~al.}(1995)\citenamefont
  {Kennett}, \citenamefont {Engdahl},\ and\ \citenamefont
  {Buland}}]{10.1111/j.1365-246X.1995.tb03540.x}%
  \BibitemOpen
  \bibfield  {author} {\bibinfo {author} {\bibfnamefont {B.~L.~N.}\
  \bibnamefont {Kennett}}, \bibinfo {author} {\bibfnamefont {E.~R.}\
  \bibnamefont {Engdahl}}, \ and\ \bibinfo {author} {\bibfnamefont
  {R.}~\bibnamefont {Buland}},\ }\href {\doibase
  10.1111/j.1365-246X.1995.tb03540.x} {\bibfield  {journal} {\bibinfo
  {journal} {Geophysical Journal International}\ }\textbf {\bibinfo {volume}
  {122}},\ \bibinfo {pages} {108} (\bibinfo {year} {1995})},\ \Eprint
  {http://arxiv.org/abs/https://academic.oup.com/gji/article-pdf/122/1/108/1543667/122-1-108.pdf}
  {https://academic.oup.com/gji/article-pdf/122/1/108/1543667/122-1-108.pdf}
  \BibitemShut {NoStop}%
\bibitem [{\citenamefont {Moser}\ and\ \citenamefont
  {V{\"o}lgyesi}(1982)}]{Moser1982THEIS}%
  \BibitemOpen
  \bibfield  {author} {\bibinfo {author} {\bibfnamefont {M.}~\bibnamefont
  {Moser}}\ and\ \bibinfo {author} {\bibfnamefont {L.}~\bibnamefont
  {V{\"o}lgyesi}},\ }\href {https://pp.bme.hu/ch/article/view/2955} {\bibfield
  {journal} {\bibinfo  {journal} {Periodica Polytechnica Chemical Engineering}\
  }\textbf {\bibinfo {volume} {26}},\ \bibinfo {pages} {155} (\bibinfo {year}
  {1982})}\BibitemShut {NoStop}%
\bibitem [{\citenamefont {Axani}(2019)}]{Axani:2019sbk}%
  \BibitemOpen
  \bibfield  {author} {\bibinfo {author} {\bibfnamefont {S.~N.~G.}\
  \bibnamefont {Axani}},\ }\emph {\bibinfo {title} {{Sterile Neutrino Searches
  at the IceCube Neutrino Observatory}}},\ \href@noop {} {Ph.D. thesis},\
  \bibinfo  {school} {MIT} (\bibinfo {year} {2019}),\ \Eprint
  {http://arxiv.org/abs/2003.02796} {arXiv:2003.02796 [hep-ex]} \BibitemShut
  {NoStop}%
\bibitem [{\citenamefont {Aartsen}\ \emph {et~al.}(2013)\citenamefont {Aartsen}
  \emph {et~al.}}]{IceCube:2013llx}%
  \BibitemOpen
  \bibfield  {author} {\bibinfo {author} {\bibfnamefont {M.~G.}\ \bibnamefont
  {Aartsen}} \emph {et~al.} (\bibinfo {collaboration} {IceCube}),\ }\href
  {\doibase 10.1016/j.nima.2013.01.054} {\bibfield  {journal} {\bibinfo
  {journal} {Nucl. Instrum. Meth. A}\ }\textbf {\bibinfo {volume} {711}},\
  \bibinfo {pages} {73} (\bibinfo {year} {2013})},\ \Eprint
  {http://arxiv.org/abs/1301.5361} {arXiv:1301.5361 [astro-ph.IM]} \BibitemShut
  {NoStop}%
\bibitem [{\citenamefont {Aartsen}\ \emph {et~al.}(2019)\citenamefont {Aartsen}
  \emph {et~al.}}]{IceCube:2019lxi}%
  \BibitemOpen
  \bibfield  {author} {\bibinfo {author} {\bibfnamefont {M.~G.}\ \bibnamefont
  {Aartsen}} \emph {et~al.} (\bibinfo {collaboration} {IceCube}),\ }\href
  {\doibase 10.1088/1475-7516/2019/10/048} {\bibfield  {journal} {\bibinfo
  {journal} {JCAP}\ }\textbf {\bibinfo {volume} {10}},\ \bibinfo {pages} {048}
  (\bibinfo {year} {2019})},\ \Eprint {http://arxiv.org/abs/1909.01530}
  {arXiv:1909.01530 [hep-ex]} \BibitemShut {NoStop}%
\bibitem [{\citenamefont {Abbasi}\ \emph
  {et~al.}(2024{\natexlab{a}})\citenamefont {Abbasi} \emph
  {et~al.}}]{IceCube:2024qxf}%
  \BibitemOpen
  \bibfield  {author} {\bibinfo {author} {\bibfnamefont {R.}~\bibnamefont
  {Abbasi}} \emph {et~al.} (\bibinfo {collaboration} {IceCube}),\ }\href
  {\doibase 10.5194/tc-18-75-2024} {\bibfield  {journal} {\bibinfo  {journal}
  {The Cryosphere}\ }\textbf {\bibinfo {volume} {18}},\ \bibinfo {pages} {75}
  (\bibinfo {year} {2024}{\natexlab{a}})}\BibitemShut {NoStop}%
\bibitem [{\citenamefont {Arg\"uelles}\ \emph {et~al.}(2022)\citenamefont
  {Arg\"uelles}, \citenamefont {Salvado},\ and\ \citenamefont
  {Weaver}}]{Arguelles:2021twb}%
  \BibitemOpen
  \bibfield  {author} {\bibinfo {author} {\bibfnamefont {C.~A.}\ \bibnamefont
  {Arg\"uelles}}, \bibinfo {author} {\bibfnamefont {J.}~\bibnamefont
  {Salvado}}, \ and\ \bibinfo {author} {\bibfnamefont {C.~N.}\ \bibnamefont
  {Weaver}},\ }\href {\doibase 10.1016/j.cpc.2022.108346} {\bibfield  {journal}
  {\bibinfo  {journal} {Comput. Phys. Commun.}\ }\textbf {\bibinfo {volume}
  {277}},\ \bibinfo {pages} {108346} (\bibinfo {year} {2022})},\ \Eprint
  {http://arxiv.org/abs/2112.13804} {arXiv:2112.13804 [hep-ph]} \BibitemShut
  {NoStop}%
\bibitem [{\citenamefont {Esmaili}\ and\ \citenamefont
  {Smirnov}(2013)}]{Esmaili:2013vza}%
  \BibitemOpen
  \bibfield  {author} {\bibinfo {author} {\bibfnamefont {A.}~\bibnamefont
  {Esmaili}}\ and\ \bibinfo {author} {\bibfnamefont {A.~Y.}\ \bibnamefont
  {Smirnov}},\ }\href {\doibase 10.1007/JHEP12(2013)014} {\bibfield  {journal}
  {\bibinfo  {journal} {JHEP}\ }\textbf {\bibinfo {volume} {12}},\ \bibinfo
  {pages} {014} (\bibinfo {year} {2013})},\ \Eprint
  {http://arxiv.org/abs/1307.6824} {arXiv:1307.6824 [hep-ph]} \BibitemShut
  {NoStop}%
\bibitem [{\citenamefont {Gariazzo}(2020)}]{Gariazzo:2019xhx}%
  \BibitemOpen
  \bibfield  {author} {\bibinfo {author} {\bibfnamefont {S.}~\bibnamefont
  {Gariazzo}},\ }\href {\doibase 10.1140/epjc/s10052-020-8126-0} {\bibfield
  {journal} {\bibinfo  {journal} {Eur. Phys. J. C}\ }\textbf {\bibinfo {volume}
  {80}},\ \bibinfo {pages} {552} (\bibinfo {year} {2020})},\ \Eprint
  {http://arxiv.org/abs/1910.06646} {arXiv:1910.06646 [astro-ph.CO]}
  \BibitemShut {NoStop}%
\bibitem [{\citenamefont {Feroz}\ \emph {et~al.}(2009)\citenamefont {Feroz},
  \citenamefont {Hobson},\ and\ \citenamefont {Bridges}}]{Feroz:2008xx}%
  \BibitemOpen
  \bibfield  {author} {\bibinfo {author} {\bibfnamefont {F.}~\bibnamefont
  {Feroz}}, \bibinfo {author} {\bibfnamefont {M.~P.}\ \bibnamefont {Hobson}}, \
  and\ \bibinfo {author} {\bibfnamefont {M.}~\bibnamefont {Bridges}},\ }\href
  {\doibase 10.1111/j.1365-2966.2009.14548.x} {\bibfield  {journal} {\bibinfo
  {journal} {Mon. Not. Roy. Astron. Soc.}\ }\textbf {\bibinfo {volume} {398}},\
  \bibinfo {pages} {1601} (\bibinfo {year} {2009})},\ \Eprint
  {http://arxiv.org/abs/0809.3437} {arXiv:0809.3437 [astro-ph]} \BibitemShut
  {NoStop}%
\bibitem [{\citenamefont {Jeffreys}(1939)}]{Jeffreys:1939xee}%
  \BibitemOpen
  \bibfield  {author} {\bibinfo {author} {\bibfnamefont {H.}~\bibnamefont
  {Jeffreys}},\ }\href@noop {} {\emph {\bibinfo {title} {{The Theory of
  Probability}}}},\ Oxford Classic Texts in the Physical Sciences\ (\bibinfo
  {year} {1939})\BibitemShut {NoStop}%
\bibitem [{\citenamefont {Abbasi}\ \emph
  {et~al.}(2023{\natexlab{b}})\citenamefont {Abbasi} \emph
  {et~al.}}]{IceCube:2023qua}%
  \BibitemOpen
  \bibfield  {author} {\bibinfo {author} {\bibfnamefont {R.}~\bibnamefont
  {Abbasi}} \emph {et~al.} (\bibinfo {collaboration} {IceCube}),\ }\href
  {\doibase 10.22323/1.444.0975} {\bibfield  {journal} {\bibinfo  {journal}
  {PoS}\ }\textbf {\bibinfo {volume} {ICRC2023}},\ \bibinfo {pages} {975}
  (\bibinfo {year} {2023}{\natexlab{b}})},\ \Eprint
  {http://arxiv.org/abs/2307.13951} {arXiv:2307.13951 [astro-ph.HE]}
  \BibitemShut {NoStop}%
\bibitem [{\citenamefont {Rukhin}\ \emph {et~al.}(2010)\citenamefont {Rukhin}
  \emph {et~al.}}]{NIST}%
  \BibitemOpen
  \bibfield  {author} {\bibinfo {author} {\bibfnamefont {A.}~\bibnamefont
  {Rukhin}} \emph {et~al.},\ }\href
  {https://csrc.nist.gov/pubs/sp/800/22/r1/upd1/final} {\bibfield  {journal}
  {\bibinfo  {journal} {{NIST SP 800-22 Rev. 1}}\ } (\bibinfo {year}
  {{2010}})}\BibitemShut {NoStop}%
\bibitem [{\citenamefont {Mahn}\ \emph {et~al.}(2012)\citenamefont {Mahn} \emph
  {et~al.}}]{SciBooNE:2011qyf}%
  \BibitemOpen
  \bibfield  {author} {\bibinfo {author} {\bibfnamefont {K.~B.~M.}\
  \bibnamefont {Mahn}} \emph {et~al.} (\bibinfo {collaboration} {SciBooNE,
  MiniBooNE}),\ }\href {\doibase 10.1103/PhysRevD.85.032007} {\bibfield
  {journal} {\bibinfo  {journal} {Phys. Rev. D}\ }\textbf {\bibinfo {volume}
  {85}},\ \bibinfo {pages} {032007} (\bibinfo {year} {2012})},\ \Eprint
  {http://arxiv.org/abs/1106.5685} {arXiv:1106.5685 [hep-ex]} \BibitemShut
  {NoStop}%
\bibitem [{\citenamefont {Adamson}\ \emph {et~al.}(2019)\citenamefont {Adamson}
  \emph {et~al.}}]{MINOS:2017cae}%
  \BibitemOpen
  \bibfield  {author} {\bibinfo {author} {\bibfnamefont {P.}~\bibnamefont
  {Adamson}} \emph {et~al.} (\bibinfo {collaboration} {MINOS+}),\ }\href
  {\doibase 10.1103/PhysRevLett.122.091803} {\bibfield  {journal} {\bibinfo
  {journal} {Phys. Rev. Lett.}\ }\textbf {\bibinfo {volume} {122}},\ \bibinfo
  {pages} {091803} (\bibinfo {year} {2019})},\ \Eprint
  {http://arxiv.org/abs/1710.06488} {arXiv:1710.06488 [hep-ex]} \BibitemShut
  {NoStop}%
\bibitem [{\citenamefont {Cheng}\ \emph {et~al.}(2012)\citenamefont {Cheng}
  \emph {et~al.}}]{MiniBooNE:2012meu}%
  \BibitemOpen
  \bibfield  {author} {\bibinfo {author} {\bibfnamefont {G.}~\bibnamefont
  {Cheng}} \emph {et~al.} (\bibinfo {collaboration} {MiniBooNE, SciBooNE}),\
  }\href {\doibase 10.1103/PhysRevD.86.052009} {\bibfield  {journal} {\bibinfo
  {journal} {Phys. Rev. D}\ }\textbf {\bibinfo {volume} {86}},\ \bibinfo
  {pages} {052009} (\bibinfo {year} {2012})},\ \Eprint
  {http://arxiv.org/abs/1208.0322} {arXiv:1208.0322 [hep-ex]} \BibitemShut
  {NoStop}%
\bibitem [{\citenamefont {Abe}\ \emph {et~al.}(2015)\citenamefont {Abe} \emph
  {et~al.}}]{Super-Kamiokande:2014ndf}%
  \BibitemOpen
  \bibfield  {author} {\bibinfo {author} {\bibfnamefont {K.}~\bibnamefont
  {Abe}} \emph {et~al.} (\bibinfo {collaboration} {Super-Kamiokande}),\ }\href
  {\doibase 10.1103/PhysRevD.91.052019} {\bibfield  {journal} {\bibinfo
  {journal} {Phys. Rev. D}\ }\textbf {\bibinfo {volume} {91}},\ \bibinfo
  {pages} {052019} (\bibinfo {year} {2015})},\ \Eprint
  {http://arxiv.org/abs/1410.2008} {arXiv:1410.2008 [hep-ex]} \BibitemShut
  {NoStop}%
\bibitem [{\citenamefont {Aartsen}\ \emph
  {et~al.}(2017{\natexlab{b}})\citenamefont {Aartsen} \emph
  {et~al.}}]{IceCube:2017ivd}%
  \BibitemOpen
  \bibfield  {author} {\bibinfo {author} {\bibfnamefont {M.~G.}\ \bibnamefont
  {Aartsen}} \emph {et~al.} (\bibinfo {collaboration} {IceCube}),\ }\href
  {\doibase 10.1103/PhysRevD.95.112002} {\bibfield  {journal} {\bibinfo
  {journal} {Phys. Rev. D}\ }\textbf {\bibinfo {volume} {95}},\ \bibinfo
  {pages} {112002} (\bibinfo {year} {2017}{\natexlab{b}})},\ \Eprint
  {http://arxiv.org/abs/1702.05160} {arXiv:1702.05160 [hep-ex]} \BibitemShut
  {NoStop}%
\bibitem [{\citenamefont {Abbasi}\ \emph
  {et~al.}(2024{\natexlab{b}})\citenamefont {Abbasi} \emph
  {et~al.}}]{Abbasi:2024ktc}%
  \BibitemOpen
  \bibfield  {author} {\bibinfo {author} {\bibfnamefont {R.}~\bibnamefont
  {Abbasi}} \emph {et~al.},\ }\href@noop {} {\  (\bibinfo {year}
  {2024}{\natexlab{b}})},\ \Eprint {http://arxiv.org/abs/2407.01314}
  {arXiv:2407.01314 [hep-ex]} \BibitemShut {NoStop}%
\end{thebibliography}%

\end{document}